\documentclass[a4paper,fleqn]{cas-dc}

\usepackage[numbers,sort&compress]{natbib}

\usepackage{tikz}
\usetikzlibrary{matrix,fit,positioning,calc,arrows.meta}

\usepackage{amssymb}
\usepackage{amsmath}
\usepackage{gensymb}
\usepackage{hyperref}
\usepackage[T1]{fontenc}
\usepackage{listings}

\usepackage{xcolor}
\usepackage{ulem}
\usepackage{soul}

\usepackage{algorithm}
\usepackage[noend]{algpseudocode}
\usepackage{subcaption}

\usepackage[english]{babel}

\graphicspath{{./figures/}{./scripts/}}

\newcommand{\jt}[1]{{#1}}

\begin{document}
\let\WriteBookmarks\relax
\def\floatpagepagefraction{1}
\def\textpagefraction{.001}

\shorttitle{foap4: AMR with OpenACC\jt{/OpenMP}, MPI, and p4est}

\shortauthors{Teunissen et al.}

\title [mode = title]{foap4: adaptive mesh refinement with OpenACC\jt{/OpenMP}, MPI, and p4est}



%

\author[1,2]{Jannis Teunissen}[
orcid=0000-0003-0811-5091
]

\cormark[1]


\ead{jannis.teunissen@cwi.nl}


\credit{Conceptualization of this study, Funding acquisition, Methodology, Software, Visualization, Writing – original draft}

\affiliation[1]{
  organization={Centrum Wiskunde \& Informatica (CWI)},
  addressline={Science Park 123},
  city={Amsterdam},
  postcode={1098 XG},
  country={The Netherlands}}

\affiliation[2]{
  organization={Centre for mathematical Plasma Astrophysics, Department of Mathematics, KU Leuven},
  addressline={Celestijnenlaan 200B},
  city={Leuven},
  postcode={3001},
  country={Belgium}}

\affiliation[3]{
  organization={Departamento de Matemática da Universidade de Aveiro and Centre for Research and Development in Mathematics and Applications (CIDMA)},
  addressline={Campus de Santiago},
  city={Aveiro},
  postcode={3810-193},
  country={Portugal}}

\affiliation[4]{
  organization={Netherlands eScience Center},
  addressline={Science Park 402},
  city={Amsterdam},
  postcode={1098 XH},
  country={The Netherlands}}

\affiliation[5]{
  organization={School of Astronomy and Space Science and Key Laboratory of Modern Astronomy and Astrophysics, Nanjing University},
  city={Nanjing},
  postcode={210023},
  country={China}}

\affiliation[6]{
  organization={Anton Pannekoek Institute for Astronomy, University of Amsterdam},
  addressline={Science Park 904},
  city={Amsterdam},
  postcode={1098 XH},
  country={The Netherlands}}

\affiliation[7]{
  organization={School of Physics and Astronomy, Yunnan University},
  city={Kunming},
  postcode={650500},
  country={China}}

\author[3]{Héctor R. Olivares Sánchez}[
orcid=0000-0001-6833-7580
]
\credit{Writing – review and editing}

\author[2]{Jesse Vos}[
orcid=0000-0003-3349-7394
]
\credit{Funding acquisition, Writing – review and editing}

\author[4]{Leon Oostrum}[
orcid=0000-0001-8724-8372
]
\credit{Writing – review and editing}

\author[4]{Johan Hidding}[
orcid=0000-0002-7550-1796
]
\credit{Writing – review and editing}

\author[4]{Victor Azizi}[
orcid=0000-0003-3535-8320
]
\credit{Writing – review and editing}

\author[5]{Yuhao Zhou}[
orcid=0000-0002-4391-393X
]
\credit{Writing – review and editing}

\author[5]{Hao Wu}[
orcid=0009-0006-9249-8468
]
\credit{Writing – review and editing}

\author[2]{Adrian Kelly}[
orcid=0009-0006-8524-008X
]
\credit{Writing – review and editing}

\author[2]{Olaf Willocx}[
orcid=0009-0002-9718-352X
]
\credit{Writing – review and editing}

\author[7]{Chun Xia}[
orcid=0000-0002-7153-4304
]
\credit{Writing – review and editing}

\author[2]{Rony Keppens}[
orcid=0000-0003-3544-2733
]
\credit{Writing – review and editing}

\author[6]{Oliver Porth}[
orcid=0000-0002-4584-2557
]
\credit{Conceptualization of this study, Funding acquisition, Methodology, Writing – review and editing}





\cortext[1]{Corresponding author}




\begin{highlights}
  \item foap4 achieves high performance on GPUs with AMR blocks of size $8^3$, $16^3$ and $32^3$.
  \item Ghost cells are updated efficiently by storing a pattern for filling and communicating ghost cells, and then executing that pattern on GPUs.
  \item The code demonstrates that a combination of OpenACC \jt{or OpenMP} with MPI can result in good performance on both GPUs and CPUs.
\end{highlights}

\begin{abstract}
  GPUs and other accelerators are increasingly used for scientific computing.
  We want to add GPU support to \jt{existing} parallel adaptive mesh refinement (AMR) codes written in Fortran.
  To understand which changes are necessary to obtain good performance we have developed foap4, an AMR framework implemented in Fortran that uses OpenACC\jt{/OpenMP}, MPI, and the p4est library.
  We discuss the design and implementation of the framework.
  Several benchmark problems are considered, in which Euler's equations of gas dynamics are solved using explicit time integration.
  These benchmarks are performed in both 2D and 3D, using static and adaptive meshes, for varying problem sizes on different hardware.
  Our results show that AMR simulations can be carried out efficiently on GPUs, even when using relatively small grid blocks of $8^3$ or $16^3$ cells.
  \jt{With NVIDIA's \texttt{nvfortran} compiler similar performance is obtained with OpenACC and OpenMP offloading. Initial results with the \texttt{gfortran} compiler and OpenACC demonstrate the portability of foap4.}
\end{abstract}


\maketitle

\section{Introduction}
\label{sec:introduction}

\begin{figure}
  \centering
  \includegraphics[width=\linewidth]{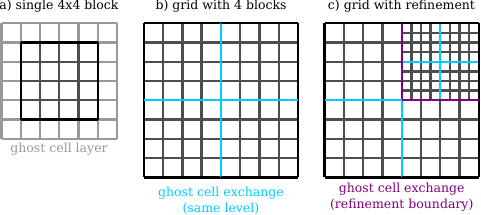}
  \caption{Schematic illustration of a quadtree mesh, with blocks containing $4\times4$ cells and a single layer of ghost cells (shaded gray).
    To fill the ghost cells around a block data has to be exchanged with neighboring blocks, which can be at the same level (light blue) or a different refinement level (purple), and which might reside on a different processor or GPU.
    The figure depicts a very simple AMR mesh; typical simulations use at least two ghost cells, larger blocks and multiple refinement levels.
  }
  \label{fig:quadtree-example}
\end{figure}

Modern supercomputers derive most of their performance and energy efficiency from GPU-equipped nodes, but many computational physics codes cannot yet run on GPUs (graphics processing units).
We would like to add GPU support to two existing parallel adaptive mesh refinement (AMR) codes written in Fortran, namely MPI-AMRVAC~\cite{keppensMPIAMRVAC30Updates2023,keppensMPIAMRVACParallelGridadaptive2021,xiaMPIAMRVAC20Solar2018} and the afivo framework~\cite{teunissenSimulatingStreamerDischarges2017,teunissenAfivoFrameworkQuadtree2018}.
\jt{These codes can be used to solve partial differential equations (PDEs) on adaptively refined meshes.}
\jt{They use quadtree (2D) / octree (3D) type AMR}, as illustrated in figure~\ref{fig:quadtree-example}.
An octree mesh consists of a collection of blocks that each have $N_x \times N_y \times N_z$ cells, but whose grid spacing can differ by a factor $2^{-l}$, where $l$ is the refinement level.

Each block is enlarged with $N_\mathrm{ghost}$ layers of ghost cells, in which part of the data of adjacent blocks is stored.
\jt{
  These ghost cells allow standard numerical operations to be performed on the interior of a block without having to take the mesh structure into account.
  The number of ghost layers required thus depends on the width of the largest numerical stencil that is used.
  For example, computing a second-order Laplacian involving only the direct neighbors requires $N_\mathrm{ghost} \geq 1$, while computing finite-volume fluxes with a second order accurate slope-limited scheme typically requires $N_\mathrm{ghost} \geq 2$.
}

The cost of AMR simulations mainly depends on two factors: how fast computations can be performed on the blocks, and how fast ghost cell data can be communicated between blocks.
This will depend on e.g., the block size, the number of ghost cells, the numerical algorithm, and the used hardware.
Also relevant is the cost for adapting the mesh and subsequent parallel load-balancing, for which we use the scalable algorithms from the p4est library~\cite{bursteddeP4estScalableAlgorithms2011}.

There are several programming models for general-purpose GPU computing~\cite{marowkaPerformancePortabilityOpenACC2022}.
To avoid a complete rewrite of our codes, we here consider a directive-based approach using \jt{either OpenACC or OpenMP}.
\jt{These directive-based approaches are} supported by different compilers and hardware, \jt{they are relatively easy to program and} they allow the same code to still run purely on CPUs.
In this paper, we describe a simple AMR code called foap4 (\textbf{F}ortran \textbf{O}pen\textbf{A}CC/OpenMP \textbf{p4}est).
We use foap4 to better understand what changes are necessary in existing AMR codes to obtain good performance on GPUs.

OpenACC has already been used in a number of computational physics codes.
Several examples are provided below, focusing on applications related to computational fluid dynamics (CFD).
In~\cite{krausAcceleratingCFDCode2014} the extension of the C++ flow solver ZFS with OpenACC was described.
The authors state that ``the directive-based programming model allows one to achieve good performance with reasonable effort, even for mature codes with many lines of code''.
They test the performance of the OpenACC implementation on a 3D flow simulation with a static mesh, using a finite volume scheme and explicit time integration.
In~\cite{caplanGPUAccelerationEstablished2019} OpenACC support was added to the solar magnetohydrodynamics (MHD) Fortran code MAS, which uses a static (but non-uniform) mesh and a combination of implicit and explicit time stepping.
The authors state that they had to modify less than 5\% of the original CPU source code for OpenACC support.

In~\cite{xueImprovedFrameworkGPU2021} the authors describe how OpenACC was added to the Fortran code SENSEI (Structured, Euler/Navier-Stokes Explicit-Implicit Solver).
The authors discuss several performance optimizations compared to an earlier OpenACC implementation~\cite{mccallMultilevelParallelismApproach2017}.
They assess the improved performance as `fair' and demonstrate the code on several 2D and 3D test cases with static grids.
\jt{Additional discussion on the porting process and the trade-offs involved in directive-based GPU programming is presented in an earlier paper~\cite{pickeringDirectivebasedGPUProgramming2015}.}
The MHD code PLUTO was recently extended with OpenACC support~\cite{rossazzaPLUTOCodeGPUs2025}, for which the authors performed a complete C++ rewrite.
The authors demonstrate the strong and weak scaling of the new gPLUTO code on several HPC systems, considering several 3D test cases on static meshes.
\jt{There are several other CFD codes in which OpenACC has already successfully been used for large-scale parallel simulations, see for example~\cite{VincentGongEtAl2022, AnChenEtAl2025, WilfongLeBerreEtAl2026}.
  It has also been demonstrated in several papers that OpenACC performance can be on par with native CUDA implementations~\cite{HoshinoMaruyamaEtAl2013,OyarzunMiraEtAl2021,ShanAraya-Polo2024}.
OpenMP offloading to accelerators appears to be less common for CFD applications, although promising results have also been obtained, see e.g.~\cite{salvadoreOpenMPOffloadExascale2024,tsoutsanisHeterogeneousExascaleCFD2026}.}

There are several parallel AMR codes and frameworks that use programming models different from OpenACC\jt{ or OpenMP} to run on GPUs.
A general block-structured AMR framework written in C++ is AMReX~\cite{zhangAMReXFrameworkBlockstructured2019}.
AMReX internally uses CUDA, HIP or SYCL for GPUs but also supports OpenACC and OpenMP device offloading in application codes.
Another general framework for block-structured AMR is Parthenon, which is written in C++ and uses the Kokkos programming model~\cite{trottKokkosEcoSystemComprehensive2021}.
There also exist several AMR application codes that use Kokkos, for example the C++ AthenaK~\cite{stoneAthenaK2026} and IDEFIX~\cite{lesurIDEFIXVersatilePerformanceportable2023} codes for astrophysical MHD simulations.
Furthermore, there are astrophysical AMR codes using CUDA, such as GAMER-2~\cite{schiveGamer2GPUacceleratedAdaptive2018} and H-AMR~\cite{liskaHAMRNewGPUaccelerated2022}.
\jt{Finally, we note that the recently developed kalypsso code~\cite{kestenerKalypssoPerformancePortable2026} shares some of the features of foap4, in particular because it also uses p4est for AMR, but a major difference is that kalypsso is based on Kokkos.}

Although there is already quite some existing work regarding parallel AMR on GPUs, it is not always clear how to extend an existing AMR Fortran code with OpenACC \jt{or OpenMP} while balancing development effort and performance.
The foap4 code presented in this paper can help in that regard:
by providing a performance reference point, better informed decisions can be made about which changes to existing codes are worth the effort.
Although it was developed for benchmarking purposes, the foap4 code can also be used to run real applications on large HPC systems.
Finally, even for applications without AMR, it might be useful to have a tool to benchmark OpenACC kernels with a simple domain-decomposition \jt{approach}.

Below, we first discuss the design and implementation of foap4 in section~\ref{sec:foap4-description}.
Several included numerical algorithms and physics modules are discussed in section~\ref{sec:included-num-phys}.
Test cases are described in section~\ref{sec:code-tests}, and performance results are presented in section~\ref{sec:performance-test}.

\section{Foap4 design and implementation}
\label{sec:foap4-description}

\subsection{Overview and limitations}
\label{sec:overview-limitations}

As described in the introduction, the mesh in foap4 is of the quadtree (2D) or octree (3D) type.
To keep the code relatively simple, several restrictions are imposed on this mesh.
\jt{First, }only Cartesian 2D/3D geometries are supported.
\jt{Second, for simplicity we do not yet include support for corner or edge ghost cells.
  Such ghost cells are necessary to handle mixed-derivative terms, which for example occur in anisotropic thermal conduction, and they are necessary in constrained-transport MHD schemes~\cite{gardinerUnsplitGodunovMethod2005}.
  For the test problems considered here we use finite volume schemes for which the numerical stencil is a `cross', i.e., no diagonal information is needed.
}
\jt{Third, } grid blocks should have the same number of cells in each dimension (e.g.\ $16 \times 16 \times 16$). This simplifies the ghost cell exchange, but the restriction could be removed relatively easily.

The overall architecture of foap4 is illustrated in figure~\ref{fig:foap4-arch}.
For the main public methods of foap4, which are prefixed by \texttt{f4\_}, the figure also shows what is performed on CPUs and on GPUs.
\jt{To aid understanding, we briefly mention a few key points below.
  Whenever the mesh refinement changes, a `ghost cell pattern' is stored that for every mesh block describes how data should be exchanged with other blocks, possibly on different MPI ranks.
  This pattern is stored in such a way that it can efficiently be executed on GPUs, as discussed in section~\ref{sec:ghost-cell-exchange}.
  Time integration can be performed with explicit Runge-Kutta schemes, as discussed in section~\ref{sec:time-integration}.
  The execution of a single Runge-Kutta step typically involves the filling of ghost cells, followed by for example a flux computation and an update of the PDE solution.
}
More details about the implementation \jt{of foap4} are provided in the following sections.


\begin{figure*}
  \centering
  \begin{tikzpicture}[
      box/.style={draw, rounded corners, align=center, inner sep=4pt},
      line/.style={-Latex, thick},
      lab/.style={midway, fill=white, inner sep=1pt}
      ]

      \matrix[row sep=4mm, column sep=8mm] {
        \node[box] (cpu-refine) {Adjust mesh (optional load balance) and\\
          store ghost cell pattern.}; &
        \node[box] (f4-refine) {\texttt{f4\_adjust\_refinement}}; &
        \node[box] (gpu-refine) {Temporarily copy block data, then\\perform interpolation and restriction}; \\
        \node[box] (cpu-update-gc) {Call GPU kernels and \\ synchronize MPI communication}; &
        \node[box] (f4-update-gc) {\texttt{f4\_update\_ghostcells}}; &
        \node[box] (gpu-update-gc) {Execute stored ghost cell pattern: \\
        fill MPI buffers and ghost cells}; \\
        \node[box] (cpu-advance) {Set Runge--Kutta scheme and\\
        call GPU kernels for each stage}; &
        \node[box] (f4-advance) {\texttt{f4\_advance}}; &
        \node[box] (gpu-advance) {Execute a Runge--Kutta stage, \\
          including e.g.\ ghost cell update, \\
          flux computation and solution update}; \\
        \node[box] (cpu-io) {Write XDMF output\\(one file per MPI rank)}; &
        \node[box] (f4-io) {\texttt{f4\_write\_grid}}; &
        \node[box] (gpu-io) {Copy block data to CPU}; \\
      };

      \node[box, fit=(cpu-refine)(cpu-update-gc)(cpu-advance)(cpu-io), inner sep=6pt] (cpuFit) {};
      \node[box, fit=(gpu-refine)(gpu-update-gc)(gpu-advance)(gpu-io), inner sep=6pt] (gpuFit) {};

      \node[box,above=2mm of cpuFit, text width=5.9cm, fill=gray!20] (cpu-main) {\textbf{CPU}\\ Manages mesh structure and connectivity (p4est with MPI)};
      \node[box,above=2mm of gpuFit, text width=5.4cm, fill=gray!20] (gpu-main) {\textbf{GPU}\\ Stores and operates on block data $\mathrm{uu}(i,j,k,i_\mathrm{var},i_\mathrm{block})$};

      \draw[line, -] (f4-refine) -- (cpu-refine);
      \draw[line, -] (f4-refine) -- (gpu-refine);

      \draw[line, -] (f4-update-gc) -- (cpu-update-gc);
      \draw[line, -] (f4-update-gc) -- (gpu-update-gc);

      \draw[line, -] (f4-advance) -- (cpu-advance);
      \draw[line, -] (f4-advance) -- (gpu-advance);

      \draw[line, -] (f4-io) -- (cpu-io);
      \draw[line, -] (f4-io) -- (gpu-io);

  \end{tikzpicture}%
  \caption{Overview of foap4 architecture with main public routines.
    Normally, one MPI rank is used per GPU.
    The GPU part can also be executed on a regular CPU.}
  \label{fig:foap4-arch}
\end{figure*}

\subsection{Mesh management with p4est}
\label{sec:mesh-management-with}

The mesh structure is managed on CPUs using p4est~\cite{bursteddeP4estScalableAlgorithms2011}.
The p4est library provides scalable algorithms for MPI-parallel AMR on a collection of connected octrees (3D) or quadtrees (2D), which are called \textit{forests} in p4est terminology.
A key feature of the library is that it does not duplicate the global mesh structure on each MPI rank.
This allows applications to scale to hundreds of thousands of cores.
By delegating mesh management to p4est, we avoid re-implementing complex algorithms for mesh adaptation, 2:1 balance constraints (ensuring neighboring blocks differ by at most one refinement level), and parallel load balancing.

\begin{figure}
  \centering
  \includegraphics[width=\linewidth]{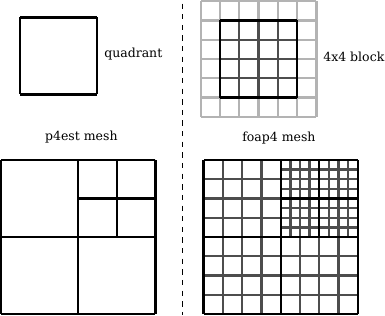}
  \caption{Illustration of the difference between the p4est mesh and the foap4 mesh. Every quadrant in p4est (a single cell) corresponds to a block of cells (including ghost cells) in foap4.}
  \label{fig:p4est-vs-foap4-block}
\end{figure}

In p4est an octree contains \textit{quadrants}, which are single grid cells.
\jt{Technically, these should be referred to as \textit{octants} in 3D, but we will simply use the term quadrants.}
In foap4 no actual data is stored on these quadrants, but we instead link every quadrant to a grid block stored on the GPU, as illustrated in figure~\ref{fig:p4est-vs-foap4-block}.
We use a simple indexing scheme: the $n$\textsuperscript{th} local quadrant in p4est (cumulative over all octrees) corresponds to the $n$\textsuperscript{th} grid block.
In a typical application, each block could for example contain $16^3$ cells and store $N_\mathrm{var}$ variables.
Every quadrant then corresponds to $16^3$ times $N_\mathrm{var}$ unknowns on the GPU, which reduces the overhead of the p4est mesh management.

Since p4est is written in C, we have written a small wrapper library in a few hundred lines of C code.
This allows the following p4est functionality to be accessed from Fortran:
\begin{itemize}
  \item Initialization and destruction of the main p4est data structures
  \item The creation of an initial `brick' geometry, consisting of a rectangular block of $N_x \times N_y \times N_z$ octrees (where $N_x$, $N_y$ and $N_z$ do not need to be equal)
  \item Obtaining the refinement levels and coordinates of local blocks (i.e., of a single MPI rank)
  \item Obtaining the connectivity of all the faces of local blocks
  \item Adjusting the mesh refinement
  \item Re-partitioning the mesh, i.e., load balancing
\end{itemize}

To perform mesh refinement, refinement flags are first determined for every local grid block.
This is typically performed on GPUs, see section~\ref{sec:amr-criterion}.
These refinement flags can have three values: positive, meaning that a block should be refined; negative, meaning that a block can be derefined, or zero, indicating that a block should at least maintain its current refinement level.
The p4est mesh is then adapted by sequentially calling \texttt{p4est\_refine}, \texttt{p4est\_coarsen} and \texttt{p4est\_balance}, with only the latter requiring parallel communication.
In the balance step we use the \texttt{P4EST\_CONNECT\_FULL} connectivity, so that blocks sharing a corner differ by at most one refinement level.
This can reduce numerical discretization errors and it simplifies the filling of ghost cells, as described in section~\ref{sec:ghost-cell-exchange}.

A convenient feature of p4est is that it keeps track of a mesh revision number, which only updates when there is an actual change in the mesh.
In case of a new mesh revision a temporary copy of the old blocks is first made on the GPU\@.
Afterwards, a loop is performed over the new blocks, comparing their refinement level to the old value.
In this way three lists are generated, containing the indices of the blocks that were refined, coarsened or unaltered.
Afterwards, the necessary prolongation (interpolation), restriction (coarsening) and copying is performed on GPUs.

Restriction is implemented by setting the coarse value to the average of the corresponding fine values.
Prolongation is performed using a limited slope determined at the cell center.
In 2D, prolongation at a coarse cell with index $(i, j)$ requires values from the neighboring cells at $(i\pm 1, j)$ and $(i, j\pm 1)$.
First, one-sided differences to the neighboring cells are determined, e.g.\ $u_{i, j} - u_{i-1, j}$ and $u_{i+1, j} - u_{i, j}$ in the $x$-direction.
A limiter is then applied to these differences to obtain $(\delta u_x, \delta u_y)$, after which the fine cells are set to $u_{i,j} + (\pm 1/4, \pm 1/4) \cdot (\delta u_x, \delta u_y)$, with the signs depending on the relative position of the fine cells.
In foap4, we use the generalized minmod limiter $G(a, b, \theta)$ for prolongation, where $a$ and $b$ are slopes and $\theta$ is a parameter.
If the signs of $a$ and $b$ differ $G(a, b, \theta) = 0$, otherwise it is given by
\begin{equation}
  \label{eq:gminmod}
  G(a, b, \theta) = \mathrm{sign}(a) \min\left(|\theta a|, |\theta b|, |a + b|/2\right).
\end{equation}
By default, we use $\theta = 1$, in which case $G$ reduces to the conventional minmod limiter.
Note that the maximum values that preserve non-negativity on a Cartesian grid with constant spacing are $\theta = 2$ in 2D and $\theta = 4/3$ in 3D, see appendix~\ref{sec:appendix-gminmod}.

Since mesh refinement can lead to an uneven distribution of the mesh over MPI ranks, partitioning (load-balancing) can be performed after refinement.
This is controlled by a threshold $\kappa_\mathrm{max}$ for the load-imbalance $\kappa$, defined as
\begin{equation}
  \label{eq:load-imbalance}
  \kappa = N_{\max} / \lceil N_{\mathrm{total}} / N_\mathrm{tasks} \rceil,
\end{equation}
where $N_{\max}$ is the maximum number of blocks on any rank, $N_{\mathrm{total}}$ is the total number of blocks across all ranks, $N_\mathrm{tasks}$ is the number of MPI ranks and $\lceil \cdot \rceil$ denotes the ceiling function.
Note that equation~\eqref{eq:load-imbalance} gives $\kappa = 1$ for a perfectly balanced distribution and $\kappa > 1$ for imbalanced distributions.
For the AMR simulations presented in this paper we use a threshold $\kappa_\mathrm{max} = 1.1$.

When partitioning is performed, a copy of the blocks is first made on each GPU.
Afterwards, partitioning is performed on the p4est mesh.
This changes the quadrant index range belonging to each MPI rank.
The index range of blocks received from and sent to other GPUs is then determined, after which a single MPI message is used between each pair of ranks that have to exchange blocks.
The MPI calls are performed using a standard floating point datatype and GPU device pointers.

\subsection{Block data structure and memory layout}
\label{sec:block-data-structure}

The primary data structure in foap4 is the grid block, which stores cell-centered solution variables.
All block data is stored in a single $(N_\mathrm{dim}+2)$-dimensional array
\begin{equation}
  \label{eq:array-structure}
  \mathrm{uu}(i, j, k, i_\mathrm{var}, i_\mathrm{block}),
\end{equation}
where the $i, j, k$ are spatial indexes, $i_\mathrm{var}$ is the variable index, and $i_\mathrm{block}$ is the block index.
Note that arrays are stored in column-major order in Fortran, so that $i$ is the fastest varying index.
Early tests showed that permuting $i_\mathrm{var}$ and $i_\mathrm{block}$ did not significantly affect performance for typical use cases.

Each block has the same number of cells in every direction ($N_x = N_y = N_z$), extended by $N_\mathrm{ghost}$ layers of ghost cells on each side.
Valid spatial indices $i, j, k$ thus range from $1 - N_\mathrm{ghost}$ to $N_x + N_\mathrm{ghost}$.
Furthermore, $N_x$ must be an even number, so that $2^{N_\mathrm{dim}}$ fine cells can be mapped to a single coarse cell when coarsening, and $N_x$ cannot be smaller than $N_\mathrm{ghost}$, the number of ghost cells used.

The refinement level and spatial origin of all local blocks are stored in arrays.
For MPI communication a single send buffer and a single receive buffer are allocated, which are simple one-dimensional arrays of floating point type.
How data is packed in these buffers is discussed in section~\ref{sec:ghost-cell-exchange}.
All the above arrays are allocated on both CPU and GPU using OpenACC \texttt{enter data} directives \jt{or their OpenMP equivalents}.
The block data array $\mathrm{uu}$ is \jt{allocated once}, with a user-defined maximum number of blocks $N_\mathrm{max blocks}$.
However, the actual array can be larger to provide storage for Runge-Kutta time integration, as described in section~\ref{sec:time-integration}.
We use static allocation because of its simplicity and because dynamically resizing the array can be problematic for problems that use almost all available memory.

By default, the block data array uses 64-bit floating point numbers, but this can be changed to 32-bit by setting a compilation flag \texttt{FLOAT\_BITS=32}.
This flag also controls the type of several other arrays, namely communication buffers, buffers for storing fluxes at refinement boundaries, and buffers for storing spatially varying boundary conditions.

\subsection{Boundary conditions}
\label{sec:boundary-conditions}

The following types of boundary conditions are supported: periodic, Dirichlet and Neumann.
Periodic boundary conditions can be specified when initializing the computational domain.
A Dirichlet boundary condition with a uniform value $u = u_b$ on the whole domain boundary can be stored per face direction and per variable in the foap4 data structure, and the same can be done for a Neumann boundary condition with a uniform normal derivative $\partial_n u = u'_b$.
(There are $2 \, N_\mathrm{dim}$ face directions, which we index in the following order: pointing to $-x$, $+x$, $-y$, $+y$, $-z$, $+z$.)

For spatially varying boundary conditions, an array \texttt{bc\_data} in the foap4 data structure can be filled.
In 3D, this array is indexed as
\begin{equation}
  \label{eq:bc-data}
  \mathrm{bc}_\mathrm{data}(i, j, i_\mathrm{var}, i_\mathrm{bf}) = \textrm{boundary value}.
\end{equation}
The index $i_\mathrm{bf}$ runs from one to the number of faces adjacent to a physical boundary and $i, j$ are the spatial indexes on the block face.
A mapping from $i_\mathrm{block}$ and the face direction to $i_\mathrm{bf}$ is stored in the foap4 data structure.
When using the $\mathrm{bc}_\mathrm{data}$ functionality, a callback function has to be stored so that boundary data is automatically updated when the mesh changes.
For temporally varying boundary conditions, this callback function can be called by the application code whenever necessary.

The $\mathrm{bc}_\mathrm{data}$ array is dynamically allocated on both CPU and GPU, since it is difficult to estimate how many faces will be adjacent to a physical boundary.
\jt{OpenACC and OpenMP GPU offloading do} not support function pointers inside loops, so the callback function mentioned above should itself contain a loop to set all the physical boundary data on the GPU.

\subsection{\jt{OpenACC and OpenMP parallelization}}
\label{sec:openacc-openmp}

\jt{We only use parallel constructs that can easily be translated between OpenACC and OpenMP.
For these constructs we define macros with the Fypp Fortran preprocessor~\cite{fypp}, which are translated to the respective offloading model depending on the compilation flags.
Table~\ref{tab:acc-omp} lists some of the main constructs and their translations.
To keep the implementation portable, loops executed on a GPU are fully collapsed when possible.
A typical loop construct is shown in figure~\ref{fig:parallel-style}.}

\begin{table*}
  \centering
  \small
  \begin{tabular}{llll}
    \hline
    Macro & OpenACC & OpenMP (GPU) & OpenMP (CPU) \\
    \hline
    \texttt{DECLARE\_DEVICE}       & \texttt{acc declare create} & \texttt{omp declare target} & - \\
    \texttt{ENTER\_DATA\_COPYIN}   & \texttt{acc enter data copyin} & \texttt{omp target enter data map(to:)} & - \\
    \texttt{ENTER\_DATA\_CREATE}   & \texttt{acc enter data create} & \texttt{omp target enter data map(alloc:)} & - \\
    \texttt{EXIT\_DATA\_DELETE}    & \texttt{acc exit data delete} & \texttt{omp target exit data map(delete:)} & - \\
    \texttt{UPDATE\_DEVICE}        & \texttt{acc update device} & \texttt{omp target update to} & - \\
    \texttt{UPDATE\_SELF}          & \texttt{acc update self} & \texttt{omp target update from} & - \\
    \hline
    \texttt{PARALLEL}              & \texttt{acc parallel} & \texttt{omp target teams} & \texttt{omp parallel}\\
    \texttt{PARALLEL\_LOOP\_FLAT}  & \texttt{acc parallel loop independent} & \texttt{omp target teams distribute parallel do} & \texttt{omp parallel do}\\
    \texttt{LOOP\_FLAT} & \texttt{acc loop independent} & \texttt{omp distribute parallel do} & \texttt{omp do}\\
    \texttt{LOOP\_OUTER}           & \texttt{acc loop independent} & \texttt{omp distribute} & \texttt{omp do} \\
    \texttt{LOOP\_INNER}           & \texttt{acc loop independent} & \texttt{omp parallel do} & -\\

    \hline
  \end{tabular}
  \caption{\jt{The main parallel macros that we define with the Fypp preprocessor and their OpenACC/OpenMP translations.
      In foap4, we essentially only use the \texttt{LOOP\_FLAT} and \texttt{PARALLEL\_LOOP\_FLAT} constructs for collapsed loops.
      Note that with OpenMP GPU offloading, there is a difference between outer and inner loops of a parallel construct, and that with OpenACC we assert to the compiler that all loops are data-independent.
      With OpenMP on the CPU nested loops are disabled to avoid oversubscription.
      Most macros accept an argument that can be used to specify e.g.\ private variables or collapse clauses.}}
  \label{tab:acc-omp}
\end{table*}

\begin{figure}
\begin{lstlisting}
${PARALLEL_LOOP_FLAT('collapse(4)')}$
do n = 1, f4%n_blocks
  do k = 1, bx(3)
    do j = 1, bx(2)
      do i = 1, bx(1)
        ! Do work per grid cell
        ...
      end do
    end do
  end do
end do
\end{lstlisting}
\caption{\jt{Example of a typical collapsed loop used in foap4. The \texttt{PARALLEL\_LOOP\_FLAT} macro is expanded according to table~\ref{tab:acc-omp}.
  Here \texttt{bx} is the block size.}}
\label{fig:parallel-style}
\end{figure}

\subsection{Ghost cell exchange}
\label{sec:ghost-cell-exchange}

For parallel AMR, ghost cells have to be exchanged between the faces of neighboring blocks.
As mentioned before, we for simplicity do not consider corner (and edge) ghost cells, but only those corresponding to direct neighbors.
Furthermore, blocks sharing a corner differ by at most one level, as discussed in section~\ref{sec:mesh-management-with}.
For a block local to one MPI rank, at refinement level $l$, and a given face direction, the following cases can occur:
\begin{itemize}
  \item SRL (same refinement level): neighbor at level $l$
  \item C2F (coarse to fine): neighbor at level $l+1$
  \item F2C (fine to coarse): neighbor at level $l-1$
  \item BC: physical boundary, see section~\ref{sec:boundary-conditions}
\end{itemize}
Furthermore, neighbors can either be local (on the same MPI rank) or non-local.

At refinement boundaries restriction and interpolation have to be performed to obtain ghost cell data at level $l-1$ and $l+1$, respectively.
We use the same restriction and limited interpolation as when changing the AMR mesh, see section~\ref{sec:mesh-management-with}.
For interpolation in a cell with index $(i, j, k)$ values at the neighbors at $i\pm 1$, $j \pm 1$ and $k \pm 1$ are required.
Since some of these values might themselves correspond to ghost cells, the filling of ghost cells is performed in two rounds. The first round (R1) is:
\begin{enumerate}
  \item Pack buffers for non-local {SRL} and {F2C} neighbors
  \item Exchange the buffers
  \item Fill ghost cells for all {SRL} and {BC} \jt{neighbors, and for the coarse side of refinement boundaries}
\end{enumerate}
The second round (R2) is:
\begin{enumerate}
  \setcounter{enumi}{3}
  \item Pack buffers for non-local {C2F} neighbors, by interpolation
  \item Exchange the buffers
  \item Fill ghost cells for \jt{the fine side of refinement boundaries}
\end{enumerate}

We implement the filling of ghost cells on GPUs almost without branching statements.
This is done by first obtaining all face connectivity information from p4est, describing the type of each face boundary, its face direction, the neighboring MPI rank, the local and remote quadrant indices, and offsets for hanging faces in case of refinement boundaries, as explained below.
From this data we construct index arrays for all the local and non-local SRL, C2F and F2C neighbors and BC faces.
These lists are sorted by the direction of the face connecting two blocks, in the order of pointing to $-x$, $+x$, $-y$, $+y$, $-z$, $+z$.
For example, all local SRL neighbors are stored so that \texttt{gc\_srl\_local(1:2, n)} are the block indices $(i_\mathrm{block}, j_\mathrm{block})$ that are connected by the $n$\textsuperscript{th} local SRL face.
Additional index arrays with a suffix \texttt{\_iface} are stored so that we can loop over the faces in a particular direction as illustrated in figure~\ref{fig:ghostcell-srl}.

For non-local SRL neighbors, \texttt{gc\_srl\_from\_buf(1:2, n)} holds $(i_\mathrm{block}, i_\mathrm{recvbuf})$, where $i_\mathrm{block}$ is the index of the local block and $i_\mathrm{recvbuf}$ is the index of the received data in the receive buffer.
Similarly, \texttt{gc\_srl\_to\_buf(1:2, n)} holds $(i_\mathrm{block}, i_\mathrm{sendbuf})$, where $i_\mathrm{sendbuf}$ is the index of the data to be sent in the send buffer.
For F2C and C2F faces we follow the same approach, but an additional offset for hanging faces is stored.
This is necessary since multiple blocks on the fine side are connected to a single block on the coarse side, namely two in 2D and four in 3D\@.
The offset indicates the relative position of a fine block within such a group.

Each MPI rank has a single send and a single receive buffer, which is used for the communication with all other ranks.
The index arrays ensure that the data in these buffers is ordered according to the following properties (from left to right): the MPI rank of the communication partner, the type of neighbor, the face direction, the block index of the receiving side, and the offsets of hanging faces (in case of refinement boundaries).
We use quicksort and counting sort to construct the index arrays.
Counting sort is used for the MPI rank of the communication partner and the type of neighbor, and quicksort is used to sort by face direction, the block index of the receiving side, and the offsets of hanging faces.

All the index arrays are computed on the CPU, but afterwards transferred to GPU memory.
The ghost cell operations are implemented using parallel loops, with separate loops for each boundary type and each face direction.
In this way, we have eliminated almost all branch statements from the ghost cell update routine.
There is only one conditional statement, namely whether the number of ghost cells is odd for a C2F boundary, in which case only part of the interpolated data from the coarse side has to be used.
A downside of our approach is that it leads to considerable code duplication.
This is alleviated to some extent by using the Fypp Fortran preprocessor~\cite{fypp}, which allows rather flexible macro calls with arguments.

Since there is considerable overhead in launching parallel loops, \jt{we use a parallel region enclosing multiple loops for each of the stages of the ghost cell update.
  For each type of boundary and face direction, we use a collapsed loop over all blocks,} as illustrated in figure~\ref{fig:ghostcell-srl}.
\jt{Finally, we remark that instead of the sorted index arrays used here it would also be possible to use hash tables, as in the kalypsso code~\cite{kestenerKalypssoPerformancePortable2026} or communicate index lists as in BATSRUS-GPU and AGILE \cite{AnChenEtAl2025,porthAstrophysicsGPUsIntroducing2026}, and it would be interesting to compare the performance of these approaches.}

\begin{figure}
\begin{lstlisting}
!$acc parallel

! Loop over all faces in the +x direction
!$acc loop collapse(4) private(iq, jq, ivar)
do n = gc_srl_local_iface(1), &
       gc_srl_local_iface(2) - 1
  do iv = 1, n_vars
    do j = 1, bx(2)
       do i = 1, n_gc
         iq = gc_srl_local(1, n)
         jq = gc_srl_local(2, n)
         ivar = i_vars(iv)

         uu(bx(1)+i, j, ivar, iq) = &
              uu(i, j, ivar, jq)
         uu(-n_gc+i, j, ivar, jq) = &
              uu(bx(1)-n_gc+i, j, ivar, iq)
      end do
    end do
  end do
end do

! Handle other face directions
...

!$acc end parallel
\end{lstlisting}
\caption{\jt{OpenACC} code fragment illustrating how ghost cells at the same refinement level (SRL) are filled.
  The case shown corresponds to a face in the $+x$ direction, with both sides present on the same MPI rank, in 2D.
  Since both sides are present, the other side is filled as well.
  The \texttt{gc\_srl\_local} array is sorted by face direction, and the \texttt{gc\_srl\_local\_iface} allows to loop over one of these directions.
  In the above code, \texttt{bx} is the block size and \texttt{n\_gc} is the number of ghost cells.
\jt{Note that a fully collapsed loop is used.}}
\label{fig:ghostcell-srl}
\end{figure}

\subsection{Flux fixing}
\label{sec:flux-fixing}

When solving transport equations it is often important to conserve mass, momentum and/or other quantities.
However, at refinement boundaries flux computations on both sides generally give different results.
To maintain conservation a flux `fixing' procedure \texttt{f4\_fix\_c2f\_flux} is included with foap4.
We have implemented flux fixing as follows.
First, during the flux computation, fluxes at refinement boundaries are stored in a pre-allocated array \texttt{bflux}, into which foap4 provides an index map.
The solution on the whole mesh is updated using the still uncorrected fluxes.
Afterwards, the \texttt{f4\_fix\_c2f\_flux} method is called, which does the following:
\begin{enumerate}
  \item For refinement boundaries shared by two different MPI ranks, boundary fluxes from the fine side are stored in a send buffer, using a similar approach as for ghost cell exchange, see section~\ref{sec:ghost-cell-exchange}.
  \item The buffers are exchanged with the non-local coarse neighbors through MPI
  \item A loop over all coarse sides of refinement boundaries is performed, and the solution in the cells adjacent to the boundary is corrected based on the difference in the coarse-side flux and the average fine-side flux.
\end{enumerate}

\subsection{Parallel XDMF output and diagnostics}
\label{sec:parallel-xdmf-output}

Solution data can be written in the XDMF (eXtensible Data Model and Format) format.
Each MPI rank writes its local block data to a separate binary file, and then sends information (block origins, grid spacing) about the written blocks to MPI rank zero.
\jt{To speed up visualization, neighboring blocks at the same refinement level are combined into larger rectangular `super-blocks'.}
An XML file describing the global mesh structure is written by rank zero.
Advantages of this approach are \jt{that there are no external dependencies}, that writing output is fast, and that it is supported by visualization tools such as VisIt and ParaView.
  The output files can optionally contain up to $N_\mathrm{ghost}$ ghost layers, where $N_\mathrm{ghost}$ is the number of ghost layers in the solution data.
  \jt{These ghost cells are used by VisIt to avoid visualization artifacts at block boundaries.
    To fill ghost cells at corners and edges, local linear extrapolation is used.}
Besides XDMF output, foap4 can also use the p4est library to write VTK files with the mesh structure and the division of the mesh over the MPI ranks.


To measure the wall-clock time spent in different parts of the foap4 library we use \texttt{MPI\_Wtime} calls.
A routine \texttt{f4\_print\_wtime} is provided that prints the time spent in different parts of the foap4 library.
This includes initialization, ghost cell filling, mesh adaptation, partitioning, grid output, ghost cell pattern updates, MPI transfers, and flux fixing.

\subsection{Source code and preprocessing}
\label{sec:source-code-prepr}

The core foap4 library consists of two main files: a small wrapper (written in C) to access p4est from Fortran, and a Fortran module \texttt{m\_foap4}.
These files are converted to 2D and 3D versions using the standard C preprocessor and the Fypp Fortran preprocessor, using statements such as:
\begin{lstlisting}[keywords={}]
#:if NDIM == 2 ! Fypp syntax
! 2D implementation
#:elif NDIM == 3
! 3D implementation
#:endif
\end{lstlisting}
We also use Fypp macro definitions to reduce the amount of code duplication in the filling of ghost cells.
The library can be compiled once and then used for any number of solution variables, any number of ghost cells and any block size.

The main application we had in mind for foap4 is solving hyperbolic partial differential equations using a finite volume scheme and explicit time integration.
Since we cannot use function pointers \jt{with OpenACC or OpenMP}, we use \texttt{include} statements to select the type of flux scheme, the limiter, and the physics routines.
An important benefit of this approach is that the compiler knows how many physical variables will be present per cell, which greatly improves the performance of some computations.

\section{Included numerical methods and physics}
\label{sec:included-num-phys}

To facilitate testing, several numerical algorithms are included with foap4 in a \texttt{numerics} subdirectory, and several physics modules are included in a \texttt{physics} subdirectory.

\subsection{Time integration}
\label{sec:time-integration}

For time integration, a method is included that supports a restricted class of explicit Runge-Kutta methods, namely those in which every sub-step can be expressed as
\begin{equation}
  \label{eq:feuler}
  \mathbf{u}(s_\mathrm{out}) = \delta t \, \mathbf{F}\left(\mathbf{u}(s_\mathrm{deriv})\right) + \sum_{i=1}^{n_\mathrm{prev}} w_i \, \mathbf{u}(s_{\mathrm{prev}, i}).
\end{equation}
Here $\mathbf{F}$ is the time derivative of $\mathbf{u}$, $\delta t$ is the time step for the sub-step, $\mathbf{u}$ is the solution on all grid blocks, $w_i$ are weights for the previous states $s_{\mathrm{prev}, i}$, $s_\mathrm{out}$ is the output state, and $s_\mathrm{deriv}$ is the state to compute time derivatives from.
Note that equation~\eqref{eq:feuler} is more restricted than the general Shu-Osher form~\cite{gottliebStrongStabilityPreservingHighOrder2001}, which allows for multiple terms with $\mathbf{F}$ on the right-hand side.
However, many well-known Runge--Kutta schemes can still be constructed, such as the optimal strong-stability preserving (SSP) schemes with two, three and four stages~\cite{ruuthHighOrderStrongStabilityPreservingRungeKutta2004}, which are all included with foap4.

As an example, Heun's two-step method can be written as
\begin{align}
  \label{eq:heun}
  \mathbf{u}_1 &= \Delta t \, \mathbf{F}\left(\mathbf{u}_t\right) + \mathbf{u}_t,\\
  \mathbf{u}_{t+1} &= \tfrac{1}{2} \Delta t \, \mathbf{F}\left(\mathbf{u}_1\right) +
  \tfrac{1}{2} \left(\mathbf{u}_t + \mathbf{u}_1\right),
\end{align}
where $\mathbf{u}_{t}$ and $\mathbf{u}_{t+1}$ refer to the previous and new time state.
Similarly, the `classical' RK4 method can be written as
\begin{align}
  \label{eq:rk4}
  \mathbf{u}_1 &= \tfrac{1}{2} \Delta t \, \mathbf{F}\left(\mathbf{u}_t\right) + \mathbf{u}_t,\\
  \mathbf{u}_2 &= \tfrac{1}{2} \Delta t \, \mathbf{F}\left(\mathbf{u}_1\right) + \mathbf{u}_t,\\
  \mathbf{u}_3 &= \Delta t \, \mathbf{F}\left(\mathbf{u}_2\right) + \mathbf{u}_t,\\
  \mathbf{u}_{t+1} &= \tfrac{1}{6} \Delta t \, \mathbf{F}\left(\mathbf{u}_3\right) +
  \tfrac{1}{3} \left(-\mathbf{u}_t + \mathbf{u}_1 + 2 \mathbf{u}_2 + \mathbf{u}_3 \right).
\end{align}
Note that this scheme is not SSP, due to the negative weight for $\mathbf{u}_t$ in the final step.

Application code has to provide a function $\mathrm{FE}$ that performs a forward Euler step of the form of equation~\eqref{eq:feuler}.
All the actual work is done in this $\mathrm{FE}$ routine, which should thus contain \jt{parallel} code to update the solution on all the blocks, \jt{for example as in algorithm~\ref{alg:fv-scheme}}.
Our interface for $\mathrm{FE}$ expects the routine to provide a limit for the next time step, based on the current solution.
\jt{Note that a single time step is used to advance all levels, i.e., there is no sub-cycling.}

The additional block storage required for the chosen Runge-Kutta scheme is allocated upon initialization.
If $N_\mathrm{states}$ temporal states are required, and $N_\mathrm{max blocks}$ is the maximum number of blocks necessary for the problem, then $N_\mathrm{states} \times N_\mathrm{max blocks}$ blocks are allocated.
Index offsets are used to access the temporal states for the Runge-Kutta sub-steps.

\subsection{Limiters and finite volume methods}
\label{sec:finite-volume}

\begin{table}
  \centering
  \caption{\jt{Slope limiters and reconstruction schemes included with foap4.
    The number of ghost cells required is indicated, as well as the typical order of accuracy.
    The maximum order of accuracy (given between parentheses) is usually only observed for smooth solutions without extrema.}}
  \label{tab:limiters}
  \begin{tabular}{lcc}
    \hline
    Scheme & Ghost cells & Order (max) \\
    \hline
    First-order upwind & 1 & 1 (1)\\
    minmod \cite{vanleerUltimateConservativeDifference1977}   & 2 & 2 (2) \\
    MC \cite{vanleerUltimateConservativeDifference1977}       & 2 & 2 (2) \\
    van Leer \cite{vanleerUltimateConservativeDifference1977} & 2 & 2 (2) \\
    Koren \cite{korenRobustUpwindDiscretization1993}    & 2 & 2 (3) \\
    WENO3 \cite{jiangEfficientImplementationWeighted1996}    & 2 & 2 (3) \\
    WENO5 \cite{jiangEfficientImplementationWeighted1996}    & 3 & $\leq 5$ (5) \\
    \hline
  \end{tabular}
\end{table}

\jt{Several slope limiters are included with foap4, as listed in table~\ref{tab:limiters}.}
Two schemes for flux computation are included as well: the local Lax-Friedrichs (Rusanov) scheme~\cite{rusanov1961calculation} and the \jt{HLLE} scheme~\cite{hartenUpstreamDifferencingGodunovType1997,einfeldtGodunovTypeMethodsGas1988}.
Furthermore, a finite volume scheme is implemented as described in algorithm~\ref{alg:fv-scheme}.
\jt{We remark that this scheme was designed for GPUs and not for regular CPUs.
  For example, face fluxes are computed twice, once from each side, which leads to simple parallel loops on GPUs but is probably not efficient on CPUs.
  Furthermore, the scheme does not exploit the large and fast caches that most CPUs have.
}

\begin{algorithm}
  \caption{\jt{Finite volume scheme}}
  \label{alg:fv-scheme}
  \begin{enumerate}
    \item Fill the required ghost cells for all blocks
    \item Call \texttt{to\_primitive} to convert \jt{all blocks} from conservative to primitive variables. The result is stored in $\mathrm{uu}_\mathrm{prim}$, which is \jt{allocated once} on each GPU.
    \item \jt{Loop over all cells of all blocks to:}
    \begin{enumerate}
      \item Call \texttt{source\_term} to compute source terms
      \item Compute cell fluxes, and subtract their divergence from the source term
      \item Compute the CFL time step per cell and apply a $\mathrm{min}$-reduction
      \item Update the solution in the cell according to the Runge--Kutta scheme
    \end{enumerate}
    \item Apply flux fixing at refinement boundaries
    \item Do an MPI reduction to determine the global next time step
  \end{enumerate}
\end{algorithm}

We use primitive-based reconstruction at cell faces, so that a physics module should provide the following routines:
\begin{itemize}
  \item \texttt{to\_primitive}: convert variables in one cell in-place to primitive form
  \item \texttt{to\_conservative}: convert variables in one cell in-place to conservative form
  \item \texttt{get\_flux}: compute the flux (in conservative variables) from primitive variables at one cell face
  \item \texttt{get\_min\_max\_wavespeed}: get the minimum and maximum wavespeed at a cell face
  \item \texttt{source\_term}: compute the source term from primitive variables in one cell
\end{itemize}

\subsection{Mesh refinement criterion}
\label{sec:amr-criterion}

To run tests with AMR, we have included a simplified version of the refinement criterion described in~\cite{lohnerAdaptiveFiniteElement1987}.
For a cell with index $i, j, k$, the following quantity is computed
\begin{equation}
  \label{eq:lohner-simple}
  E_n = \frac{|u_{n-1} - 2 u_{n} + u_{n+1}|}
  {\epsilon_\mathrm{abs} + |u_{n} - u_{n-1}| + |u_{n+1} - u_{n}| + \epsilon u^*},
\end{equation}
where $u^* = |u_{n-1}| + 2 |u_{n}| + |u_{n+1}|$, $n = i, j, k$ (with $u_{i-1}$ denoting the neighbor in the $-i$ direction, etc.), and where $\epsilon_\mathrm{abs}$ and $\epsilon$ are constants.
Refinement is then based on the norm of $|\mathbf{E}| = \sqrt{E_i^2 + E_j^2 + E_k^2}$ and two additional constants $c_\mathrm{refine}$ and $c_\mathrm{derefine}$:
\begin{itemize}
  \item If in any cell in a block $|\mathbf{E}| > c_\mathrm{refine}$, mark the block for refinement
  \item If in all cells in a block $|\mathbf{E}| < c_\mathrm{derefine}$, mark the block for derefinement
  \item Otherwise, mark the block for keeping its current refinement level
\end{itemize}
Note that whether the blocks will actually be (de)refined also depends on the refinement flags of their neighbors.
Compared to the original implementation of equation~\eqref{eq:lohner-simple} in~\cite{lohnerAdaptiveFiniteElement1987}, which is also used in MPI-AMRVAC, we do not consider cross-derivative terms, since they require diagonal ghost cells.
Furthermore, we have added a term $\epsilon_\mathrm{abs}$ to avoid division by zero.
Additional refinement criteria can easily be implemented by writing a custom routine that sets refinement flags per grid block.

\subsection{Scalar advection}
\label{sec:scalar-advection}

For testing and benchmarking, we include a scalar advection problem
\begin{equation}
  \frac{\partial \rho}{\partial t} + \nabla \cdot (\rho \mathbf{v}) = 0,
\end{equation}
where $\mathbf{v}$ is a velocity field.
For this problem, primitive and conservative variables are identical, and the minimum and maximum wave speeds in direction $i$ are $\min(v_{i}, 0)$ and $\max(v_{i}, 0)$.



\subsection{Euler gas dynamics}
\label{sec:impl-euler-gas}

We have implemented the compressible Euler equations of gas dynamics in two or three spatial dimensions:
\begin{equation}
  \label{eq:euler}
  \frac{\partial}{\partial t}
  \begin{pmatrix}
    \rho \\[3pt]
    \rho \mathbf{u} \\[3pt]
    E
  \end{pmatrix}
  +
  \nabla \cdot
  \begin{pmatrix}
    \rho \mathbf{u} \\[3pt]
    \rho \mathbf{u}\otimes\mathbf{u} + p\,\mathbf{I} \\[3pt]
    (E + p)\mathbf{u}
  \end{pmatrix}
  = \mathbf{S},
\end{equation}
where $\rho$ is mass density, $\mathbf{u}$ is the velocity, $p$ is the pressure, and $\mathbf{S}$ is a source term.
The total energy density $E$ is given by
\begin{equation}
  E = \frac{p}{\gamma - 1} + \frac{1}{2} \rho |\mathbf{u}|^2,
\end{equation}
where $\gamma$ is the ratio of specific heats, and sound speeds are computed as
\begin{equation}
  c = \sqrt{\gamma \, p / \rho}.
\end{equation}
As discussed in section~\ref{sec:finite-volume}, we use a finite-volume discretization in which primitive variables $(\rho, \mathbf{u}, p)$ are reconstructed on cell faces.
Minimum and maximum wavespeeds at a cell face are computed using Einfeldt's approach~\cite{einfeldtGodunovTypeMethodsGas1988}, as is also used in MPI-AMRVAC.

\section{\jt{Correctness and convergence tests}}
\label{sec:code-tests}

\jt{Several code tests are described below.
We have verified that these tests run correctly with different compilers (\texttt{nvfortran} and \texttt{gfortran}) and with different offloading options (OpenACC, OpenMP or OpenMP-CPU).}

\subsection{Testing ghost cell filling, interpolation and restriction}
\label{sec:code-tests-amr}

To test AMR functionality, we use a test case in the unit square (2D) or unit cube (3D) with a linear solution
\begin{equation}
  \label{eq:refinement-test}
  \phi(x, y, z) = x + 2y + 3z.
\end{equation}
This solution is also specified on the domain boundaries using the method described in section~\ref{sec:boundary-conditions}.
The prolongation and restriction algorithms implemented in foap4 should be exact for a linear solution.
To test this, we pick a refinement coordinate
\begin{equation}
  \label{eq:refinement-coord}
  \mathbf{r}_\mathrm{ref} = (0.5 + 0.49 \, i, 0.5 + 0.49 \, j, 0.5 + 0.49 \, k),
\end{equation}
where $i, j, k$ are integers from $-1$ to $1$.
In the domain, $\mathbf{r}_\mathrm{ref}$ can thus be close to a corner, the middle of a face, the middle of an edge or the center.
A loop is then performed to add $N$ levels of refinement for the blocks containing $\mathbf{r}_\mathrm{ref}$.
Each time the mesh is refined, the ghost cells are updated and it is checked whether the solution in the domain (including ghost cells) stays within a floating point error of $10^{-15}$ compared to the analytic solution.
Afterwards, the refinement is removed one level at a time, and the same check is performed.
This test is repeated for all combinations $i, j, k$ from $-1$ to $1$ and for one to four ghost cell layers.
In 3D, up to five refinement levels are added, while in 2D up to seven levels are added.

\subsection{Scalar advection \jt{-- uniform velocity}}
\label{sec:scalar-advection-test}

To test the finite volume scheme, the advection equation from section~\ref{sec:scalar-advection} is simulated on the unit square (2D) or cube (3D) using periodic boundary conditions.
The initial condition is
\begin{equation}
  \label{eq:advection-sphere-sol}
  \rho(\mathbf{r}, t_0) =
  \begin{cases}
    1, & d < R - \delta,\\[4pt]
    1 - 3q^2 + 2q^3, & R - \delta \le d < R,\\[4pt]
    0, & d \ge R,
  \end{cases}
\end{equation}
where $d$ is the distance from the domain center, $q = 1 + (d - R)/\delta$, $R = 0.1$ and $\delta = 0.05$.
The velocity field is given by $v_i = 1$ for all dimensions $i$.
A scalar advection equation with a constant velocity has a simple analytic solution, so it is a suitable problem for testing the correct implementation of numerical schemes.
Two such tests are described below.

\subsubsection{Testing flux fixing}
\label{sec:test-flux-fixing}

\begin{figure}
  \centering
  \includegraphics[width=\linewidth]{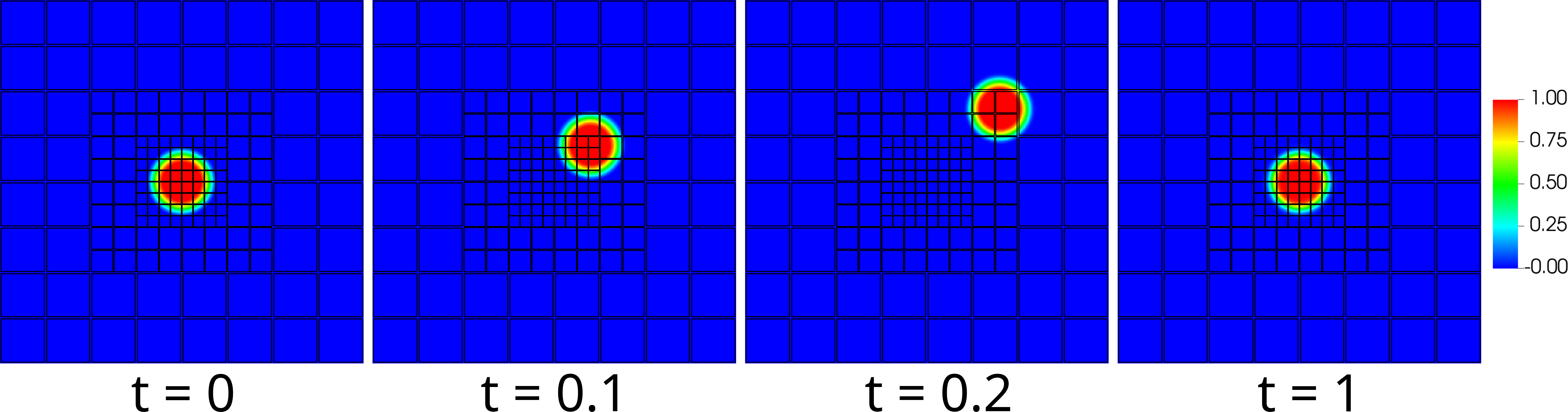}
  \caption{Illustration of test for flux fixing.
    A scalar advection equation is solved on a statically refined mesh, so that the solution crosses several refinement boundaries.
    The pictures correspond to a 2D case with blocks of size $32^2$ and five levels of initial refinement.
    Each block is indicated by a square.}
  \label{fig:advection-flux-fix-test}
\end{figure}

To test the flux fixing algorithm at refinement boundaries described in section~\ref{sec:flux-fixing}, we use static mesh refinement for the initial solution and then advance the solution up to $t = 1$.
We use a velocity field $\mathbf{v} = (1, 1)$ in 2D and $\mathbf{v} = (1, 1, 1)$ in 3D, so that at $t=1$ the solution in the unit square/cube is approximately the initial condition again.
Figure~\ref{fig:advection-flux-fix-test} shows a 2D case of this test.
The conservation error is computed as
\begin{equation}
  \label{eq:cons-error}
  e_\mathrm{cons} = \int_V \rho(\mathbf{r}, t) \, dV - \int_V \rho(\mathbf{r}, t_0) \, dV,
\end{equation}
with $V$ being the computational domain.
We have verified that $e_\mathrm{cons}$ is on the order of machine precision in 2D and 3D, on GPUs and on CPUs.

\subsubsection{\jt{Convergence test}}
\label{sec:test-convergence-uniform-vel}

In this test we compare numerical solutions of the \jt{scalar advection equation} against analytic solutions.
The initial solution given by equation~\eqref{eq:advection-sphere-sol} is not ideal for this purpose since its second derivative is discontinuous.
Therefore, we instead use a Gaussian initial condition
\begin{equation}
  \label{eq:advection-sphere-gauss}
  \rho(\mathbf{r}, t_0) = \exp(-d^2/R^2),
\end{equation}
where $d$ is the distance from the domain center and $R = 0.1$.
We again use a velocity field $\mathbf{v} = (1, 1)$ in 2D and $\mathbf{v} = (1, 1, 1)$ in 3D\jt{, and simulate up to $t = 0.1$}.

\begin{figure}
  \centering
  \includegraphics[width=\linewidth]{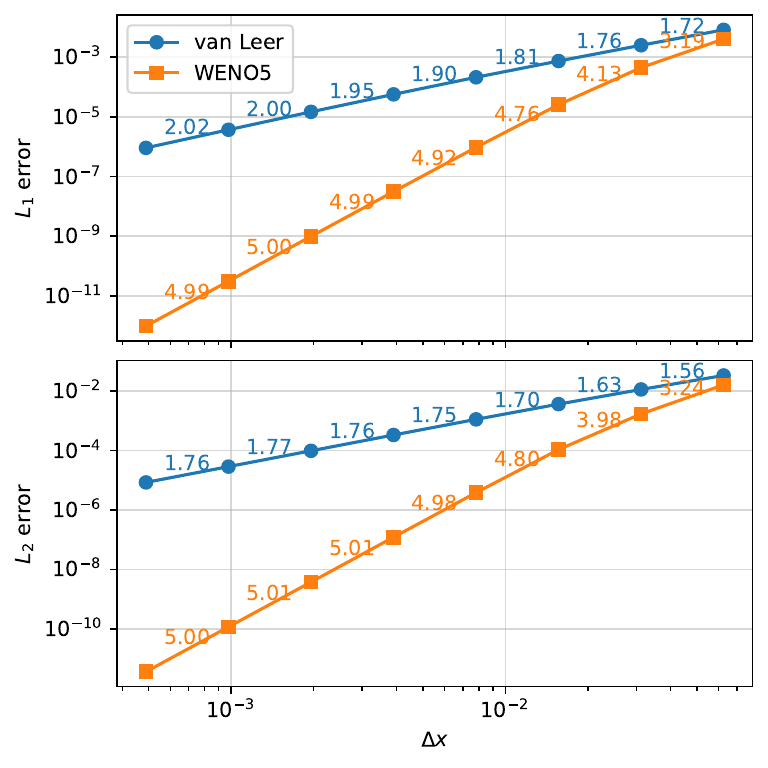}
  \caption{Convergence test results for a scalar advection equation with a Gaussian initial solution, see equation~\eqref{eq:advection-sphere-gauss}.
    Both the $L_1$ and $L_2$ errors are shown as a function of grid spacing $\Delta x$.
    The numbers on the lines indicate the estimated order of convergence between adjacent points.
    The tests were performed on uniform 2D grids with blocks of size $16^2$.}
  \label{fig:advection-convergence}
\end{figure}

Figure~\ref{fig:advection-convergence} shows $L_1$ and $L_2$ errors as a function of grid spacing $\Delta x$ using two limiters: the van Leer limiter, which is up to second order accurate, and the WENO5 limiter, which is up to fifth order accurate.
For time stepping, Heun's method was used for the van Leer case and the classic RK4 scheme for the WENO5 case.
To prevent temporal errors with the fourth-order accurate RK4 scheme from becoming important with the fifth-order WENO5 scheme, the CFL number was varied as $\Delta x^{1/4}$, with the coarsest grid corresponding to a CFL number of $0.5$.
All simulations were performed on uniform grids.

The results show the expected orders of convergence for a smooth problem.
With WENO5, the order of convergence is close to $5.0$ for sufficiently fine grids, in both error norms.
With the van Leer limiter, the $L_1$ error is close to second order, and the $L_2$ order of convergence is between $1.7$ and $1.8$ \jt{for sufficiently fine grids}.
This slight order reduction is expected due to the scheme's dissipation near extrema.

\subsection{\jt{Scalar advection -- deforming flow}}
\label{sec:scalar-advection-deform}

\begin{figure}
  \centering
  \includegraphics[width=\linewidth]{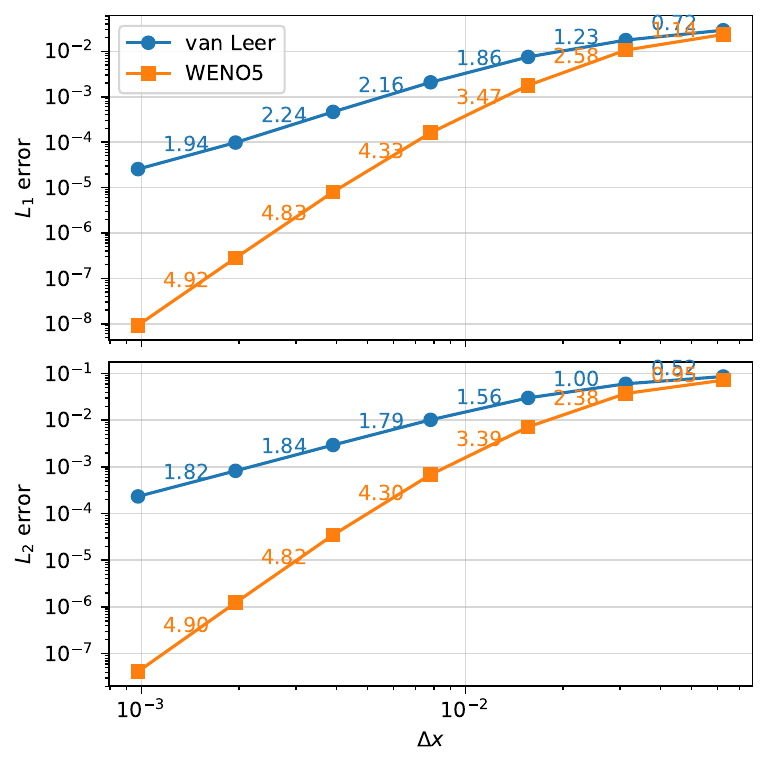}
  \caption{\jt{Convergence results for a deforming flow in 2D.
    The tests were performed on uniform 2D grids with blocks of size $16^2$.
    A Gaussian initial condition was used, given by equation~\eqref{eq:advection-sphere-gauss}.}
  }
  \label{fig:advection-convergence-deform-2d}
\end{figure}

\begin{figure}
  \centering
  \includegraphics[width=\linewidth]{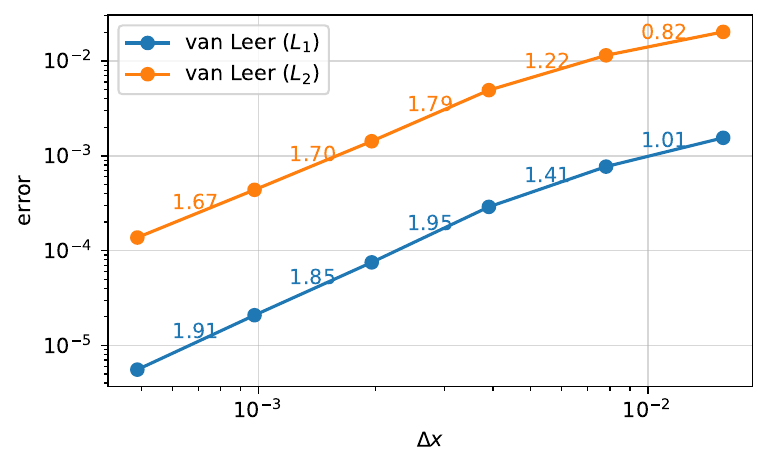}
  \caption{\jt{Convergence results for a deforming flow in 3D with AMR, using blocks of size $16^3$.
    Equation~\eqref{eq:advection-sphere-sol} is used as an initial condition, see section~\ref{sec:scalar-advection-deform} for details. $\Delta x$ indicates the spacing on the finest grid.}
  }
  \label{fig:advection-convergence-deform-3d}
\end{figure}

\begin{figure}
  \centering
  \includegraphics[width=\linewidth]{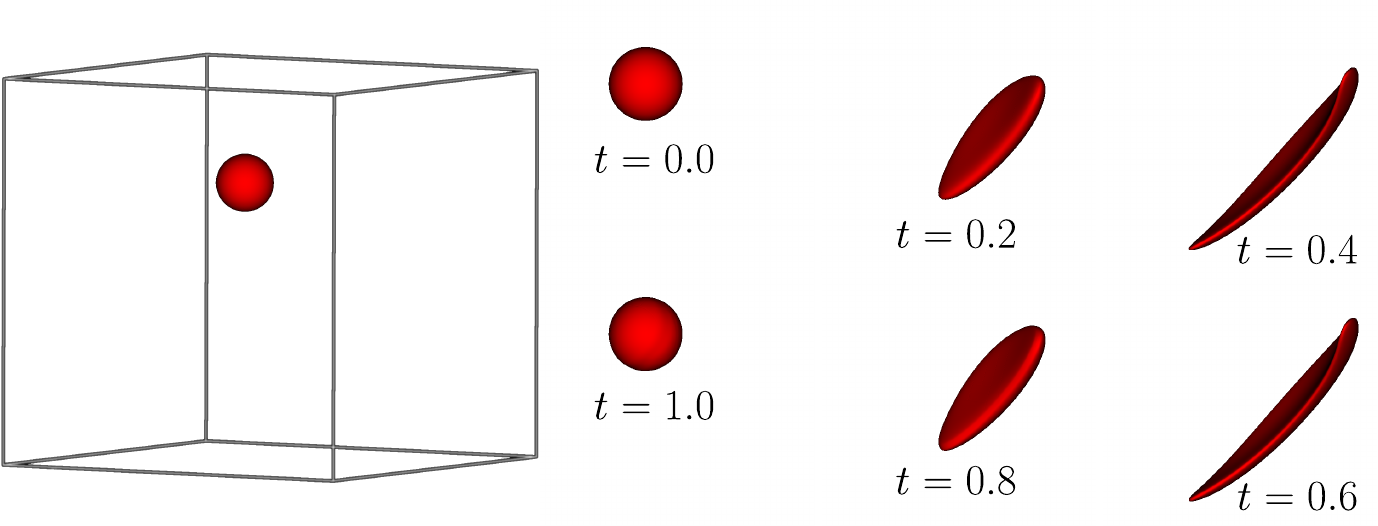}
  \caption{\jt{Solution of the 3D deforming flow problem.
    Left: the unit cube domain with the initial condition.
    Right: contour surface of $\rho = 0.5$ over time.
    The simulations were performed with six levels of refinement, so that the finest grid has a spacing of $\Delta x = 1/1024$.
    At $t = 0.5$ the flow reverses, see section~\ref{sec:scalar-advection-deform} for details.}}
  \label{fig:advection-convergence-deform}
\end{figure}

\jt{As a more challenging convergence test, we again consider the scalar advection equation but now with a velocity field that deforms the solution~\cite{LeVeque_1996}.
The velocity field in 2D is given by
\begin{equation}
  \label{eq:flow-deform-2d}
  \begin{aligned}
    v_x &= -\sin^2(\pi x)\,\sin(2\pi y)\, g(t), \\
    v_y &=  \sin(2\pi x)\,\sin^2(\pi y)\, g(t),
  \end{aligned}
\end{equation}
and in 3D by
\begin{equation}
  \label{eq:flow-deform-3d}
  \begin{aligned}
    v_x &= 2 \sin^2(\pi x)\,\sin(2\pi y)\,\sin(2\pi z)\, g(t), \\
    v_y &= -\sin(2\pi x)\,\sin^2(\pi y)\,\sin(2\pi z)\, g(t), \\
    v_z &= -\sin(2\pi x)\,\sin(2\pi y)\,\sin^2(\pi z)\, g(t),
  \end{aligned}
\end{equation}
where
\begin{equation}
  g(t) = \cos\!\left(\pi t/T\right),
\end{equation}
where we use $T = 1$.
The problem is solved on the unit square (2D) and unit cube (3D).
As described in~\cite{LeVeque_1996} this velocity field reverses at $t = T/2$, so that the analytic solution $\rho(t = T)$ is identical to the initial condition $\rho(t = 0)$.
We can thus compare these solutions to determine numerical errors.
Time steps are determined from the CFL condition, with a limit $\Delta t \leq 10^{-3}$ to avoid problems when the velocity becomes zero at $t = T/2$.

First, we do a convergence test on a uniform grid in 2D, using the velocity field from equation~\eqref{eq:flow-deform-2d}.
The same time integration as in section~\ref{sec:test-convergence-uniform-vel} is used, an initial Gaussian profile is centered at $(x, y) = (0.7, 0.7)$, see equation~\eqref{eq:advection-sphere-gauss}.
Convergence results are shown in figure~\ref{fig:advection-convergence-deform-2d}.
Both the van Leer and WENO5 scheme give the expected order of convergence on sufficiently fine grids.

Next, we consider the 3D flow field described by equation~\eqref{eq:flow-deform-3d}.
As an initial condition we use the initial solution given by equation~\eqref{eq:advection-sphere-sol}, centered at $(x, y, z) = (0.7, 0.7, 0.7)$.
The mesh is refined using $c_\mathrm{refine} = 0.1$, $c_\mathrm{derefine} = 0.0125$ and $\epsilon = 0.01$, see section~\ref{sec:amr-criterion}.
Grid blocks of $16^3$ are used, with the minimum refinement level $l = 2$ corresponding to a $64^3$ grid.
The maximum refinement level is varied from $l = 2$ to $l = 6$, with the latter corresponding to a $1024^3$ grid.
The van Leer limiter is used and Heun's method (second order) for time integration.
The numerical convergence behavior is shown in figure~\ref{fig:advection-convergence-deform-3d}, and the temporal evolution of the solution is illustrated in figure~\ref{fig:advection-convergence-deform}.
The $L_1$ error shows approximately second order behavior, while the $L_2$ error has a slightly lower convergence rate, which is probably due to non-smoothness of the initial condition (in its second derivative).
We remark that with a Gaussian initial condition and the WENO5 scheme higher-order convergence like in figure~\ref{fig:advection-convergence-deform-2d} could be obtained, but only if a large volume was refined that covered the tail of the Gaussian.}

\subsection{\jt{Blast wave test}}
\label{sec:sedov-blast-wave}

\jt{To test the Euler solver we consider the classic Sedov--von Neumann--Taylor blast wave problem, using the parameters given in~\cite{stoneAthenaAdaptiveMesh2020}.
The domain is the unit square (2D) or unit cube (3D).
Initially, the density is uniform $\rho_0 = 1$, the gas is at rest, the background pressure is $p_0 = 10^{-3}$ and the background energy density is $e_0 = p_0 / (\gamma - 1)$.
At the center of the domain, a total energy $E_\mathrm{tot} = 1$ is deposited within a spherical region with radius $R = 10^{-2}$.
An adiabatic index $\gamma = 5/3$ is used, so the pressure inside the spherical region is initially about $p = 1.6\times 10^5$ in the 3D case and about $p = 2.1\times 10^3$ in the 2D case.

To ensure that exactly $E_\mathrm{tot}$ is deposited on the numerical mesh the initial condition is specified in two steps.
First, an energy density $e_\mathrm{blast} = E_\mathrm{tot}/V_\mathrm{sphere}$ is added to all cells whose center falls within the radius $R$, where $V_\mathrm{sphere}$ is the analytic volume of the sphere.
Then the actual total deposited energy is determined by numerically integrating over the domain, from which a correction to $e_\mathrm{blast}$ is determined that ensures the total energy equals $E_\mathrm{tot}$.
Reference solutions for the problem are computed using the ExactPack code, available at \url{https://github.com/lanl/ExactPack}.
Note that in these idealized solutions the initial energy is present as a point-like source.
}

\begin{figure*}
  \centering
  \includegraphics[width=\linewidth]{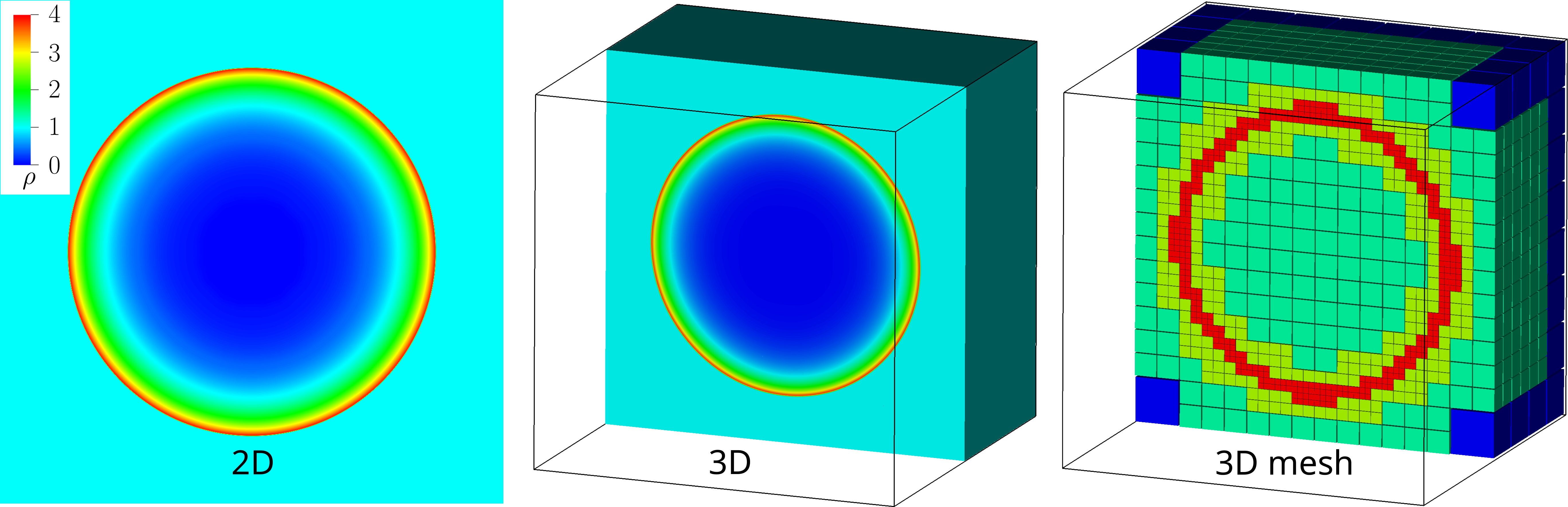}
  \caption{\jt{Density $\rho$ in 2D and 3D blast wave problems, at respectively $t = 0.1$ and $t = 0.06$.
    The 2D simulation was performed on a uniformly refined mesh with $\Delta x = 1/2048$.
    The 3D simulation used AMR and six levels of refinement, with $\Delta x = 1/1024$ on the finest grid.
    The structure of the AMR mesh is illustrated on the right, with the colors indicating the refinement level.
    Every cube corresponds to a block of $16^3$ cells.
    The simulations were performed in the unit square (2D) or cube (3D).}
  }
  \label{fig:blast-wave-sol}
\end{figure*}

\begin{figure}
  \centering
  \includegraphics[width=\linewidth]{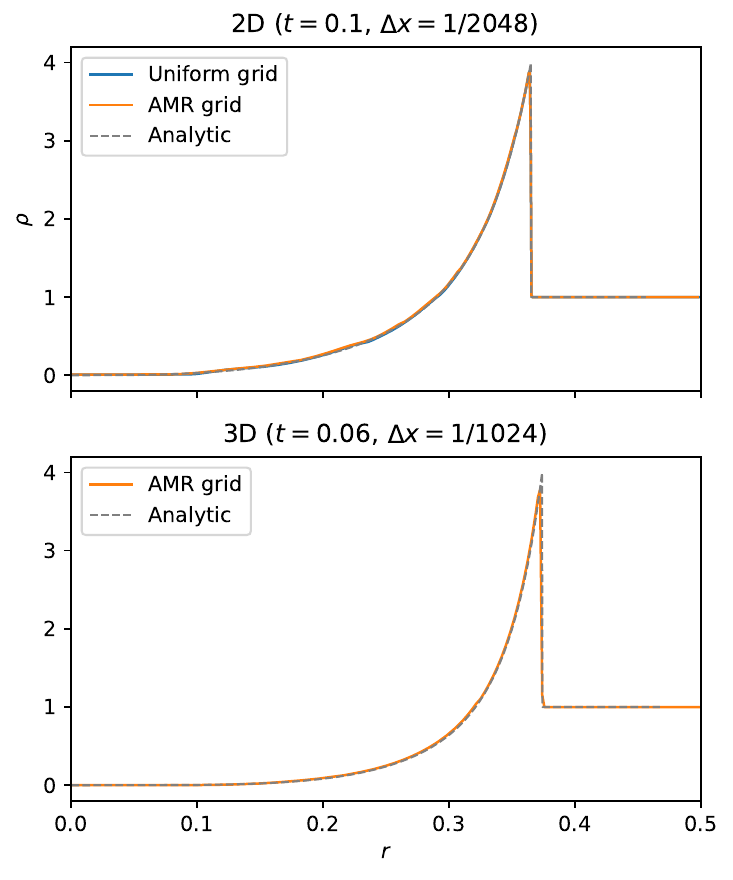}
  \caption{\jt{Comparison of 2D and 3D blast wave solutions against analytic solution computed with the ExactPack code.}
  }
  \label{fig:blast-wave-comparison}
\end{figure}

\begin{figure}
  \centering
  \includegraphics[width=\linewidth]{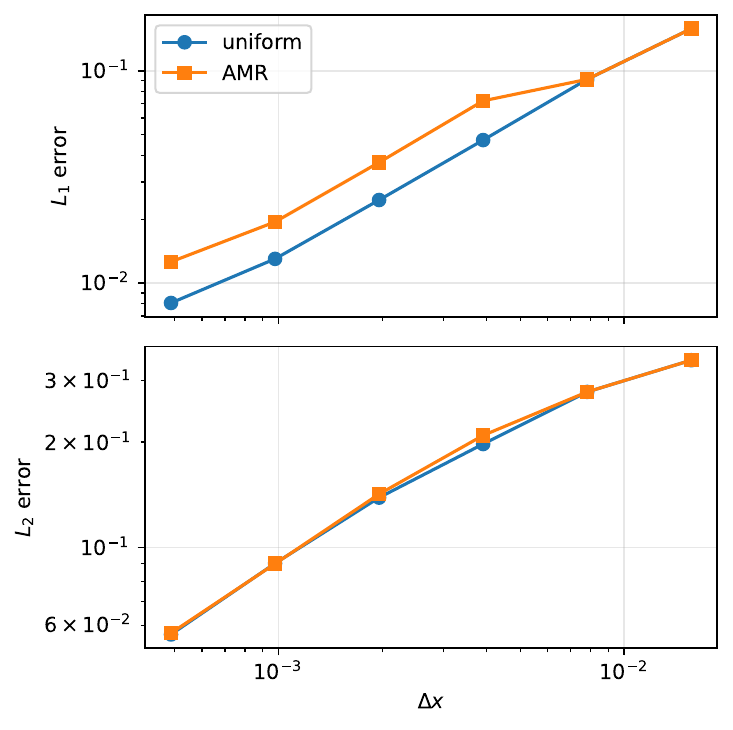}
  \caption{\jt{Convergence of 2D blast wave solutions compared to analytic solution.
    For the cases with AMR, the grid spacing refers to the finest grid.
    The errors were computed from the density, extracted along a line $x = y$, see text for details.}
  }
  \label{fig:blast-wave-convergence}
\end{figure}

\jt{We use the \jt{HLLE} scheme, the van Leer limiter and Heun's method (second order) for time integration.
For the cases with AMR, refinement is based on the energy density, using $c_\mathrm{refine} = 0.1$, $c_\mathrm{derefine} = 0.0125$ and $\epsilon = 0.01$.
In 2D, solutions are computed up to $t = 0.1$, while in 3D they are computed up to $t = 0.06$.
Note that the 2D problem corresponds to a cylindrical blast wave, and the 3D one to a spherical blast wave, hence the difference.

Figure~\ref{fig:blast-wave-sol} shows the density at the final time for the 2D and 3D solutions.
To compare with a reference solution from ExactPack, we extract the density $\rho$ along a line $x = y$ in 2D and $x = y = z$ in 3D.
Results are shown figure~\ref{fig:blast-wave-comparison}.
In all cases, the position of the shock is accurately predicted by the simulations.
For the 2D case, both AMR and uniform-grid solutions were computed, which agree well.
In 2D, small wiggles in the density profile are visible behind the shock front, which are not present in the analytic solution.

Finally, figure~\ref{fig:blast-wave-convergence} shows the $L_1$ and $L_2$ errors in the density for the 2D cases with and without AMR\@.
These were computed by interpolating the extracted numerical solutions at the analytic solution points.
Since the solutions are defined on a line, weights proportional to the radius (distance from the center) were used to obtain volume averages.
The $L_1$ error shows approximately first order convergence, while the $L_2$ error is of order less than one.
Similar behavior has been observed in previous convergence studies~\cite{timmes2005automated}.
There could be several reasons for this: the initial condition is not truly point-like (and spherical), and near the shock front our standard finite volume code reduces to first order.}

\section{Performance results}
\label{sec:performance-test}

\subsection{Rayleigh--Taylor test}
\label{sec:rt-test}

\jt{For performance testing, we simulate a Rayleigh--Taylor instability in a gravitational field, using Euler's equations as given in section~\ref{sec:impl-euler-gas}.}
The domain is the unit square/cube with periodic boundaries in the horizontal directions (i.e., $x, y$ in 3D).
An interface between a heavy and a light fluid is placed at height $h_0 = 0.8$, with densities
$\rho_\text{high} = 1.0$ above and $\rho_\text{low} = 0.1$ below the interface.
Pressure is initialized to balance a gravitational acceleration $\mathbf{g} = (0, -1)$ in 2D and $\mathbf{g} = (0, 0, -1)$ in 3D. A small sinusoidal perturbation is applied to the interface height $h$, which in 3D is given by
\begin{equation*}
  h(x, y) = h_0 + \delta \, \sin(k_1 x) \, \sin(k_2 y),
\end{equation*}
with $\delta = 0.05$, $k_1 = k_2 = 2\pi$.
In 2D, the $y$ term is absent.
All velocity components are initially zero, and the adiabatic index $\gamma$ is set to $5/3$.
Gravitational acceleration is included through a source term in equation~\eqref{eq:euler}
\begin{equation}
  \mathbf{S} =
  \begin{pmatrix}
    0 \\[3pt]
    \rho \, \mathbf{g} \\[3pt]
    \rho \, \mathbf{g} \cdot \mathbf{u}
  \end{pmatrix}.
\end{equation}

\subsection{General information}
\label{sec:gen-info}

In the performance results reported below the main metric is GCUPS, which stands for giga ($10^9$) cell updates per second.
\jt{Specifically, we measure how many interior cells (excluding ghost cells) are updated per second of total wall-clock execution time.
  A full step of an $n$-stage Runge-Kutta scheme counts as $n$ updates per cell, so that the GCUPS value is approximately} independent of the choice of Runge-Kutta scheme.
\jt{Initialization and setting up initial conditions is not included in the measured wall-clock time, and no output was written to disk during the benchmarks.}

Testing was performed on the Dutch national supercomputer Snellius.
For GPU tests we mainly used \texttt{gcn\_h100} nodes containing four NVIDIA H100 GPUs, with each H100 having 94 GiB HBM2e memory, and two AMD EPYC 9334 CPUs.
\jt{We furthermore used \texttt{gcn\_a100} nodes with four A100 GPUs, each having 40 GiB of HBM2 memory, and two Intel Xeon 8360Y CPUs.}
The foap4 code was compiled with \texttt{NVHPC 25.3}, \texttt{CUDA 12.8} and \texttt{OpenMPI 5.0.7}, using the \texttt{-fast} optimization flag.
For CPU scaling tests we used \texttt{genoa} nodes containing two AMD Genoa 9654 CPUs, for a total of 192 cores per node, and 24 banks of 16 GiB \jt{DDR5-4800} memory.
We used \texttt{gfortran 14.2}, \texttt{OpenMPI 5.0.8}, and the \texttt{-Ofast -march=native} optimization flags.
All tests were performed using 64-bit floating point numbers, except when stated otherwise.

\subsection{Uniform grid, single GPU}
\label{sec:test-uniform}

To understand how different block sizes, domain sizes and slope limiters affect performance, we first consider two test cases on uniform grids: the scalar advection test described in section~\ref{sec:scalar-advection-test} and the Rayleigh--Taylor (RT) test described in section~\ref{sec:rt-test}.
For both problems we perform time integration with Heun's method, \jt{which is }a two-step Runge-Kutta scheme, using a CFL number of $0.5$.
Results on a single NVIDIA H100 GPU are listed in table~\ref{tab:performance-uniform}.
Two limiters are used for these tests, the van Leer limiter, which requires two layers of ghost cells, and the WENO5 limiter, which requires three layers of ghost cells.

\jt{The results in table~\ref{tab:performance-uniform} show that performance differences between OpenACC and OpenMP offloading are typically within 10--20\% when using the \texttt{nvfortran} compiler, with neither approach consistently faster.
  This is certainly a positive result from a portability point of view.
  A further demonstration of portability is given in appendix~\ref{sec:appendix-gfortran}, where initial performance results for OpenACC offloading with \texttt{gfortran} are presented.}

\jt{Another important conclusion from table~\ref{tab:performance-uniform} is that although the highest performance is typically obtained with the largest block size, the drop in performance when going to small blocks is generally not severe.
  For example, with blocks of size $8^3$ we see only about a 35--40\% performance drop compared to blocks of size $64^3$ for the Euler test case.
  Since smaller blocks allow for more sparse refinement, the time-to-solution can be lower with small blocks, as was recently demonstrated in~\cite{kestenerKalypssoPerformancePortable2026}.
}

\jt{Note that GCUPS values are typically 4--10 times lower for the RT test compared to the scalar advection case, due to the larger number of variables and more expensive flux computation.
  Performance is 40\% to 60\% lower with the WENO5 limiter compared to the van Leer case, due to WENO5's larger numerical stencil (with three ghost cells) and because it requires more floating point computations.
}

\begin{table*}
\begin{tabular}{l l l l | l r | l r}
  &          &       & & \multicolumn{2}{c|}{Advection (GCUPS)} & \multicolumn{2}{c}{Euler (GCUPS)} \\
  Limiter  & domain   & block  & $N_\mathrm{blocks}$ & {OpenACC} & {OpenMP} & {OpenACC} & {OpenMP} \\
  \hline
  van Leer & 2048$^2$ & 64$^2$ & 1024    & 11.87 & 13.78 & 3.11 & 3.07 \\
  2D       &          & 32$^2$ & 4096    & 11.15 & 12.88 & 2.89 & 2.84 \\
           &          & 16$^2$ & 16384   & 9.67  & 10.95 & 2.54 & 2.47 \\
           &          & 8$^2$  & 65536   & 7.57  & 8.17  & 2.04 & 1.97 \\
           & 8192$^2$ & 64$^2$ & 16384   & 20.61 & 20.11 & 3.58 & 3.32 \\
           &          & 32$^2$ & 65536   & 18.25 & 17.86 & 3.29 & 3.06 \\
           &          & 16$^2$ & 262144  & 14.56 & 14.38 & 2.74 & 2.63 \\
           &          & 8$^2$  & 1048576 & 10.20 & 9.97  & 2.17 & 2.08 \\
  \hline
  van Leer & 256$^3$  & 64$^3$ & 64      & 10.84 & 11.71 & 1.66 & 1.55 \\
  3D       &          & 32$^3$ & 512     & 10.05 & 11.16 & 1.57 & 1.52 \\
           &          & 16$^3$ & 4096    & 8.48  & 9.18  & 1.35 & 1.31 \\
           &          & 8$^3$  & 32768   & 6.23  & 6.56  & 1.07 & 1.02 \\
           & 512$^3$  & 64$^3$ & 512     & 15.54 & 15.26 & 1.69 & 1.59 \\
           &          & 32$^3$ & 4096    & 13.62 & 13.30 & 1.59 & 1.50 \\
           &          & 16$^3$ & 32768   & 10.72 & 10.30 & 1.36 & 1.29 \\
           &          & 8$^3$  & 262144  & 7.24  & 7.17  & 1.06 & 1.01 \\
  \hline
  WENO5    & 2048$^2$ & 64$^2$ & 1024    & 6.12  & 7.30  & 1.46 & 1.48 \\
  2D       &          & 32$^2$ & 4096    & 5.81  & 6.86  & 1.36 & 1.36 \\
           &          & 16$^2$ & 16384   & 5.13  & 5.99  & 1.19 & 1.19 \\
           &          & 8$^2$  & 65536   & 4.19  & 4.66  & 1.02 & 1.01 \\
           & 8192$^2$ & 64$^2$ & 16384   & 8.10  & 8.68  & 1.49 & 1.54 \\
           &          & 32$^2$ & 65536   & 7.57  & 8.01  & 1.38 & 1.39 \\
           &          & 16$^2$ & 262144  & 6.53  & 6.82  & 1.19 & 1.21 \\
           &          & 8$^2$  & 1048576 & 5.00  & 5.19  & 0.98 & 1.03 \\
  \hline
  WENO5    & 256$^3$  & 64$^3$ & 64      & 5.48  & 6.01  & 0.71 & 0.68 \\
  3D       &          & 32$^3$ & 512     & 5.10  & 5.72  & 0.68 & 0.67 \\
           &          & 16$^3$ & 4096    & 4.39  & 4.79  & 0.61 & 0.60 \\
           &          & 8$^3$  & 32768   & 3.34  & 3.52  & 0.47 & 0.49 \\
           & 512$^3$  & 64$^3$ & 512     & 6.20  & 6.44  & 0.71 & 0.71 \\
           &          & 32$^3$ & 4096    & 5.73  & 5.88  & 0.69 & 0.68 \\
           &          & 16$^3$ & 32768   & 4.75  & 4.89  & 0.62 & 0.61 \\
           &          & 8$^3$  & 262144  & 3.52  & 3.58  & 0.48 & 0.49 \\
  \hline
\end{tabular}
\caption{Performance results \jt{with OpenACC and OpenMP GPU offloading} for the scalar advection test case and Rayleigh--Taylor test case on uniform grids, using a single H100 GPU. GCUPS stands for $10^9$ cells updated per second. The domain size, block size, and slope limiter are varied. The van Leer limiter requires two ghost cells while the WENO5 limiter requires three.}
\label{tab:performance-uniform}
\end{table*}

An H100 is a powerful GPU, but the foap4 code can also run on \jt{other} hardware.
This is illustrated in table~\ref{tab:performance-hardware}, which shows results for the Rayleigh--Taylor test case on a sample of older and newer hardware.
The tests were performed on a $256^3$ uniform grid in combination with the van Leer limiter.
We include results using both 64-bit (fp64) and 32-bit (fp32) floating point numbers for block data.

\jt{The CPU results in table~\ref{tab:performance-hardware} were obtained with OpenMP CPU threading and the \texttt{gfortran} compiler.
The $N_\mathrm{tasks}$ column contains the number of MPI tasks times the number of OpenMP threads.
For the listed combinations performance is not sensitive to the number of threads per task.
}

\jt{Memory bandwidth appears to be an important factor for performance, since the finite volume scheme with the van Leer limiter has a rather low arithmetic intensity (i.e., number of computations per memory load/store).
A simple estimate for the memory bandwidth that ignores caching is presented below.
For every cell, the stencil for the flux computation contains $(1 + 2 \, N_\mathrm{ghost})$ cells per dimension.
Furthermore, there is a conversion to primitive (one load, one store), a source term has to be added, and the final state has to be written.
The number of bytes $n_\mathrm{bytes}$ loaded or stored from memory per cell update can thus be estimated as
\begin{equation}
  \label{eq:mem-estimate}
  n_\mathrm{bytes} \approx \frac{\mathrm{bits}}{8} \, N_\mathrm{var} \, \left(3 + N_\mathrm{dim} \, (1 + 2 \, N_\mathrm{ghost})\right),
\end{equation}
where $\mathrm{bits}$ is 64 or 32, $N_\mathrm{var} = 2 + N_\mathrm{dim}$ for the Euler case and $N_\mathrm{ghost} = 2$ for the van Leer limiter.
For 64-bit numbers this gives $n_\mathrm{bytes} = 720$ and for 32-bit numbers $n_\mathrm{bytes} = 360$.
If we multiply the GCUPS values for $32^3$ blocks from table~\ref{tab:performance-hardware} with $n_\mathrm{bytes}$, the estimated bandwidth utilization on the H100 and A100 is about 50\%, while for the 1080 Ti it is about 15-30\%.
On the tested CPUs the bandwidth utilization lies in a similar range (25--50\%) for fp64, but it is lower for fp32.

Besides memory bandwidth, floating point performance is also important.
The H100 and A100 GPUs have a 2:1 ratio of theoretical fp32 and fp64 performance, while this ratio is 32:1 on a GTX 1080 Ti.
This probably explains why the relative performance of the 1080 Ti is better with fp32.
}

\begin{table*}
  \begin{tabular}{l l l l l c | l l l | l l l}
     & & & & & & \multicolumn{3}{c|}{GCUPS (fp64)} & \multicolumn{3}{c}{GCUPS (fp32)}\\
    Hardware & RAM & \jt{Bandwidth} & Year & Type & $N_\mathrm{tasks}$ & $32^3$ & $16^3$ & $8^3$ & $32^3$ & $16^3$ & $8^3$\\
    \hline
    \jt{NVIDIA H100} & 94 GiB & 2.3 TB/s & 2022 & GPU & 1 & 1.555 & 1.347 & 1.063 & 3.164 & 2.774 & 2.231\\
    \jt{NVIDIA A100} & 40 GiB & 1.4 TB/s & 2020 & GPU & 1 & 0.977 & 0.870 & 0.694 & 2.046 & 1.784 & 1.446\\
    \hline
    \jt{NVIDIA 1080 Ti} & 11 GiB & 0.5 TB/s & 2017 & GPU & 1 & 0.099 & 0.096 & 0.087 & 0.410 & 0.378 & 0.254 \\
    \hline
    \multirow{4}*{\jt{2x AMD 9654}} & \multirow{4}*{384 GiB} & \multirow{4}*{0.9 TB/s} & \multirow{4}*{2022} & \multirow{4}*{CPU} & 192$\times$1 & 0.313 & 0.312 & 0.251 & 0.355 & 0.363 & 0.318\\
     & & & & & 48$\times$4 & 0.348 & 0.327 & 0.268 & 0.377 & 0.391 & 0.341\\
     & & & & & 24$\times$8 & 0.329 & 0.326 & 0.268 & 0.408 & 0.389 & 0.337\\
     & & & & & 12$\times$16 & 0.334 & 0.312 & 0.255 & 0.396 & 0.339 & 0.315\\
    \hline
    \jt{Intel i7-10700K} & 64 GiB & 0.03 TB/s & 2020 & CPU & 8$\times$1 & 0.024 & 0.020 & 0.016 & 0.031 & 0.028 & 0.022 \\
  \hline
\end{tabular}
\caption{Performance results on different hardware for the Rayleigh--Taylor test case on a uniform $256^3$ grid with the van Leer limiter and block sizes of $32^3$, $16^3$ and $8^3$.
  The table includes results using either 64-bit or 32-bit floating point numbers for block data.
  \jt{The `bandwidth' column shows the theoretical memory bandwidth of the hardware.
    Small differences compared to table~\ref{tab:performance-uniform} are due to run-to-run variation.}
  For CPU runs we used OpenMP (CPU), and the $N_\mathrm{tasks}$ column contains the number of MPI tasks followed by the number of OpenMP threads.
}
\label{tab:performance-hardware}
\end{table*}

\jt{Our performance results can be compared with some of those reported for the AthenaK code~\cite{stoneAthenaK2026}.
  One test case in this paper is a 3D Euler gas dynamics problem solved on a $512 \times 256 \times 256$ uniform mesh and blocks of size $128^3$, using a two-step Runge-Kutta scheme, piecewise linear reconstruction, and the HLLE Riemann solver.
  It appears that the metrics given in~\cite{stoneAthenaK2026} are for a full update of a cell, and thus not for an individual Runge-Kutta substep like we have done here.
  Correcting for this, the effective GCUPS values of~\cite{stoneAthenaK2026} are 1.228 (single A100) and 1.354 (single H100).
  With blocks of size $32^3$ we get better performance on an H100 and lower performance on an A100.
  This difference could be explained by the fact that there are several versions of the A100 and H100 available, with different memory bandwidths.
}

\subsection{Parallel scaling on uniform grid}
\label{sec:test-uniform-scaling}

\begin{figure}
  \centering
  \includegraphics[width=\linewidth]{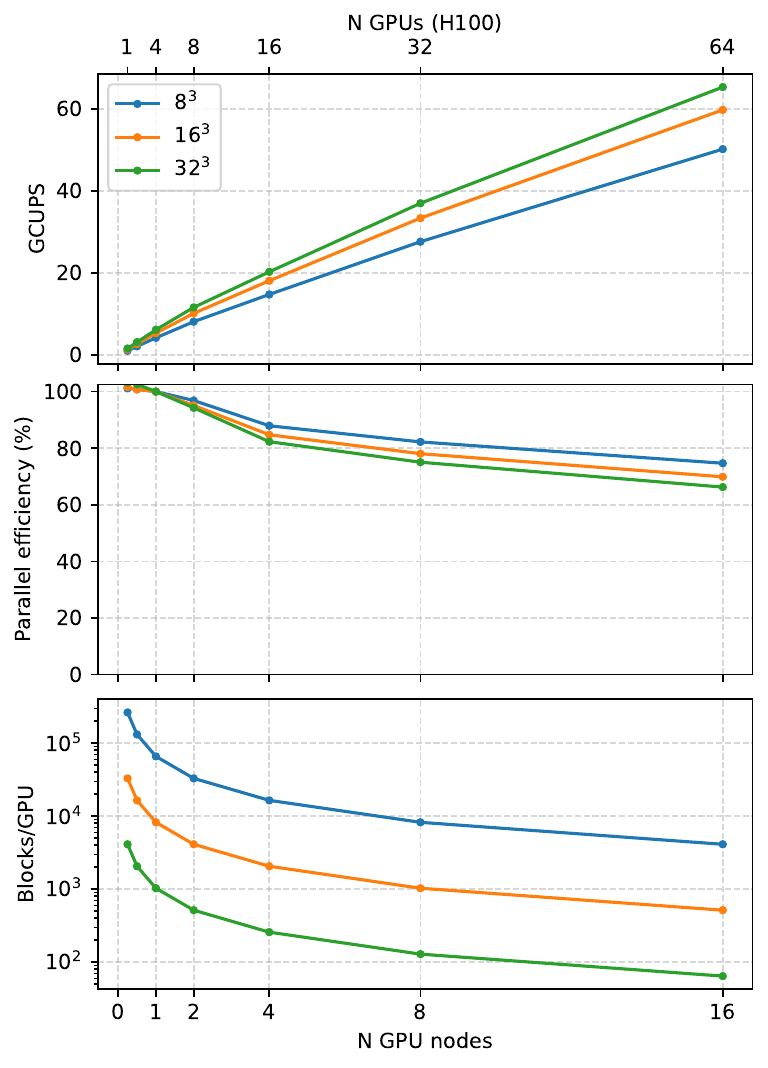}
  \caption{GPU strong scaling results for the Rayleigh--Taylor test case on a uniform $512^3$ grid \jt{with OpenACC}. Block sizes of $8^3$, $16^3$ and $32^3$ were used, as indicated in the legend. The tests were performed on nodes containing four H100 GPUs. The left-most results were performed using 1 and 2 H100s, with 1 GPU corresponding to 1/4 node. Parallel efficiency is normalized to the case using one full node.}
  \label{fig:strong-scaling-gpu}
\end{figure}

We now study parallel scaling on a uniform grid, using the 3D Rayleigh--Taylor test case with the van Leer limiter.
For these tests nodes containing four H100 GPUs were used, as described in section~\ref{sec:gen-info}, with one MPI task per GPU.
Strong scaling results on a $512^3$ grid with different block sizes are shown in figure~\ref{fig:strong-scaling-gpu}.
The parallel efficiencies in this figure are normalized to the case of a single full node.
\jt{Parallel efficiencies are about 80\% with 8 nodes and about 65\% to 75\% with 16 nodes, with the efficiencies being slightly higher for smaller blocks (since there are more of them), but GCUPS values generally being higher for larger blocks.}
With 16 nodes we get GCUPS values between \jt{50 ($8^3$ blocks) and 65 ($32^3$ blocks)}.
Since the full domain contains $512^3$ cells, an update of all the cells thus takes about \jt{two to three milliseconds} of wall-clock time.

\begin{figure}
  \centering
  \includegraphics[width=\linewidth]{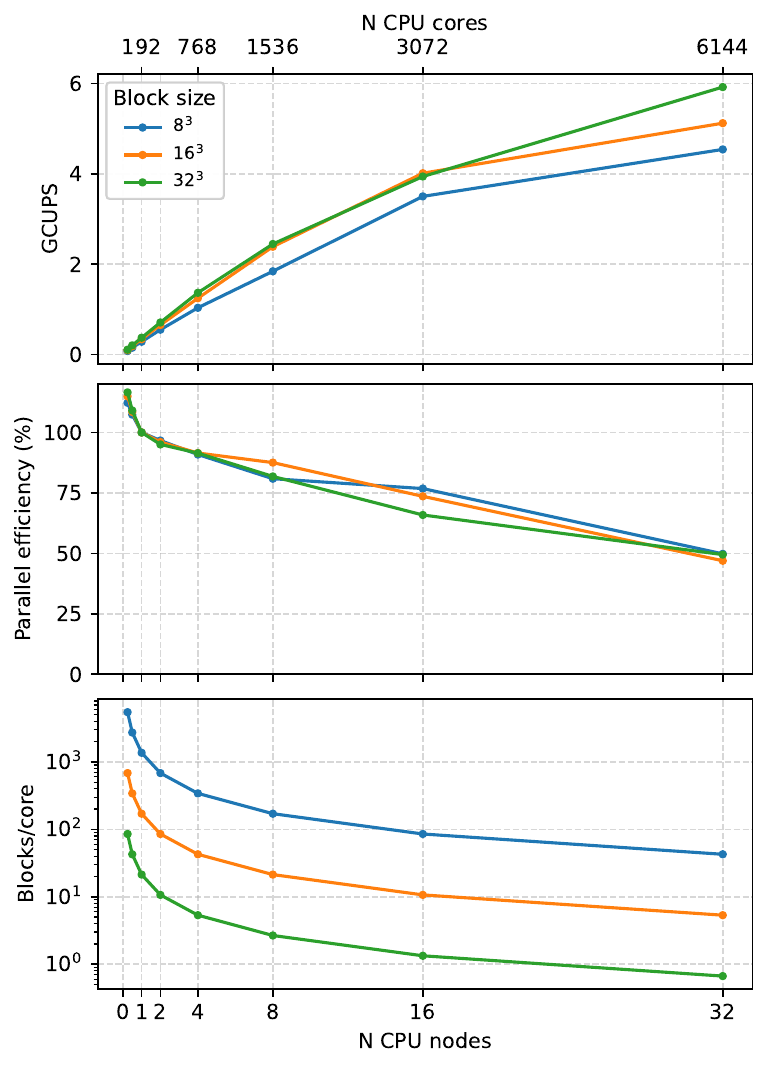}
  \caption{CPU strong scaling results for the Rayleigh--Taylor test case on a uniform $512^3$ grid.
    The tests were performed on nodes containing two AMD Genoa 9654 CPUs.
    The left-most results were performed using 48 and 96 cores.
    Parallel efficiency is normalized to the case using one full node.
    \jt{Eight OpenMP CPU threads were used per MPI task.}}
  \label{fig:strong-scaling-cpu}
\end{figure}

For comparison, strong scaling results on CPUs are shown in figure~\ref{fig:strong-scaling-cpu}.
\jt{Per MPI task eight OpenMP CPU threads were used.}
When using multiple nodes we observed a spread in GCUPS values of about $\pm$ 10\%, which was probably caused by a different node assignment and thus a different communication pattern on the HPC system.
We therefore performed three runs for each test case, and show results of the best-performing run.
With 32 nodes (and 6144 \jt{CPU cores}) parallel efficiencies \jt{are about 50\%}.


\jt{For this test case GCUPS values are about 15 times higher on GPU nodes compared to CPU nodes, with parallel scaling behaving similarly on both types of nodes.
  It is therefore possible to obtain much higher GCUPS values on GPU nodes.
  Even if the performance on CPUs was improved by say a factor two, by optimizing the finite volume kernel, there would still almost be an order-of-magnitude difference in performance.
}

To study weak scaling we consider the same test case, but now using approximately $512^3$ cells per H100 GPU\@.
A cubic domain of size $N_x^3$ is used, with $N_x$ given by
\begin{equation}
  \label{eq:weak-scaling-size}
  N_x = 4 \, \mathrm{bx} \, \lfloor N_\mathrm{GPU}^{1/3} \, \frac{512}{4 \, \mathrm{bx}} \rfloor,
\end{equation}
where $\mathrm{bx}$ is the block size and $N_\mathrm{GPU}$ the number of GPUs.
This expression ensures that the number of blocks per dimension is a multiple of four, since we noticed that an odd number could result in slightly lower performance due to a less efficient distribution of blocks over MPI ranks.

\begin{figure}
  \centering
  \includegraphics[width=\linewidth]{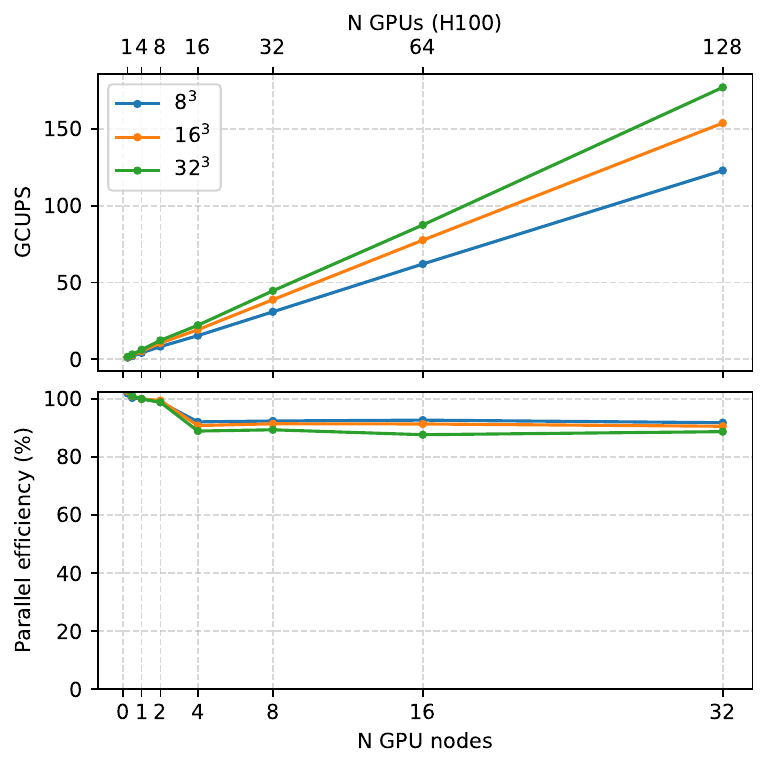}
  \caption{GPU weak scaling results for the Rayleigh--Taylor test case \jt{using OpenACC}.
    The computational \jt{domain was scaled so that approximately $512^3$ cells were} used per H100 GPU, see equation~\eqref{eq:weak-scaling-size}.
    Parallel efficiency is normalized to the case using one full node.}
  \label{fig:weak-scaling}
\end{figure}

Results are shown in figure~\ref{fig:weak-scaling}.
The parallel efficiency is about 90\% when between 4 and 32 nodes are used.
Since \jt{parallel efficiencies} are normalized to the single-node case, a scaling efficiency below 100\% is to be expected.
One reason for this is that the number of neighbors that each MPI rank has to communicate with initially increases with the number of MPI tasks.
A second reason is that communication between nodes is typically slower than communication within a single node.

\subsection{Strong scaling with AMR}
\label{sec:test-amr}

We again consider the Rayleigh--Taylor test case described in section~\ref{sec:rt-test}, but now with adaptive mesh refinement.
As a refinement criterion equation~\eqref{eq:lohner-simple} is used on the gas density, with $c_\mathrm{refine} = 5 \times 10^{-2}$, $c_\mathrm{derefine} = 6.25 \times 10^{-3}$, $\epsilon = 10^{-2}$ and $\epsilon_\mathrm{abs} = 10^{-10}$.
Refinement is performed every four time steps.
We use the van Leer limiter, a block size of $16^3$, and simulations run until $t = 1.5$.

\begin{figure}
  \centering
  \begin{subfigure}[b]{\linewidth}
    \caption{}
    \label{fig:amr-mesh-time-a}
    \includegraphics[width=\linewidth]{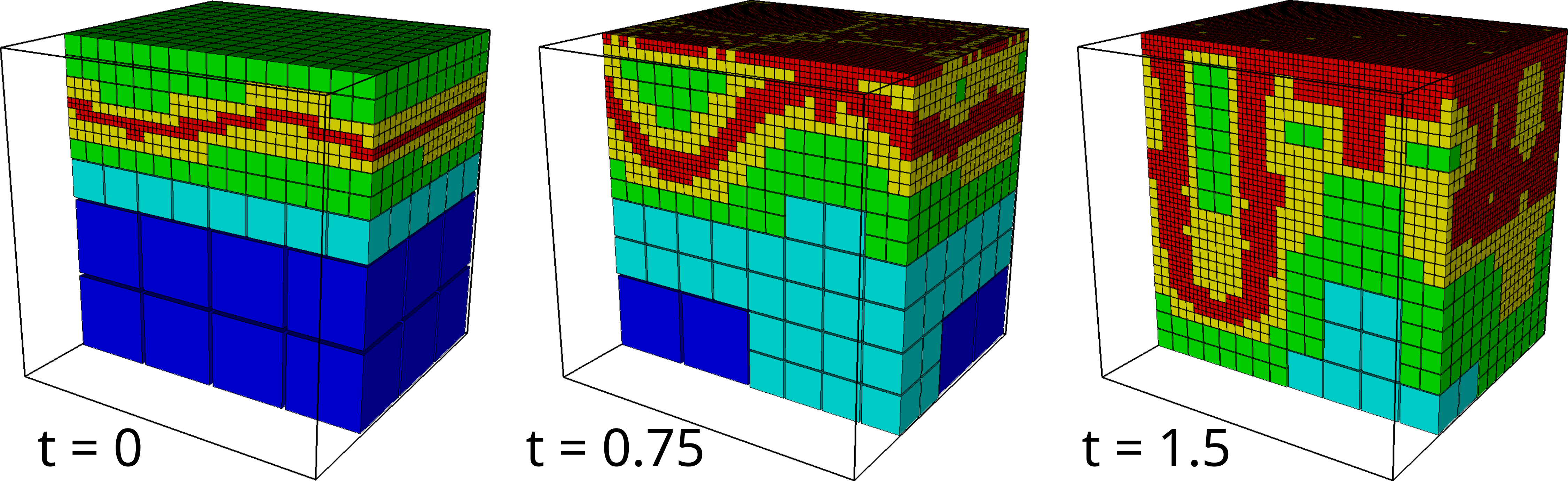}
  \end{subfigure}
  \begin{subfigure}[b]{\linewidth}
    \caption{}
    \label{fig:amr-mesh-time-b}
    \includegraphics[width=\linewidth]{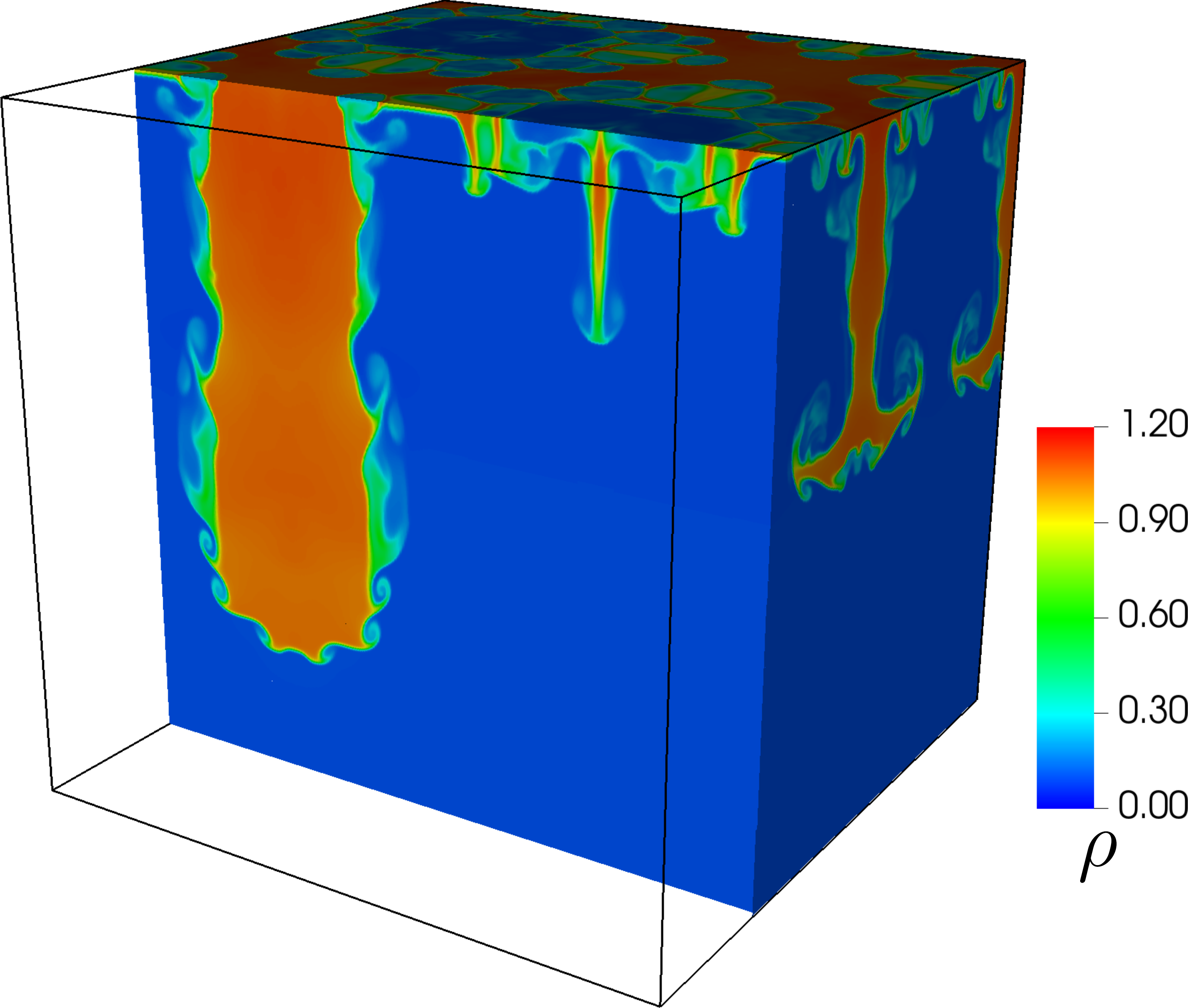}
  \end{subfigure}
  \begin{subfigure}[b]{\linewidth}
    \caption{}
    \label{fig:amr-mesh-time-c}
    \includegraphics[width=\linewidth]{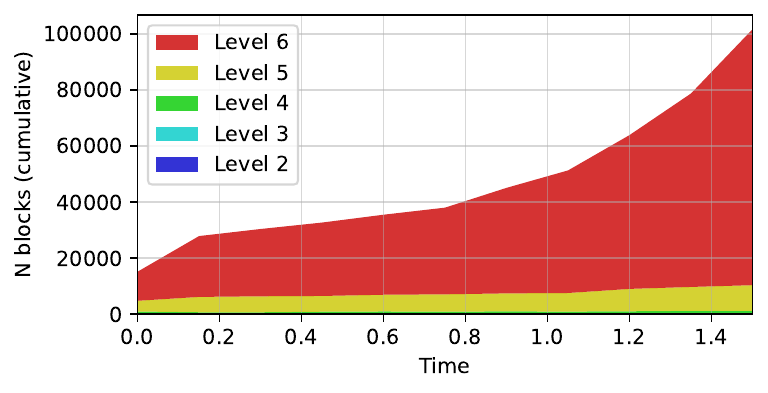}
  \end{subfigure}
  \caption{(a) Evolution of the mesh over time in the Rayleigh--Taylor test case with AMR.
    The colors correspond to refinement levels 2 (blue) to 6 (red).
    Each cell corresponds to a $16^3$ block.
    Part of the computational domain, which is the unit cube, is cut out.
    (b) Density $\rho$ at $t = 1.5$, rendered with VisIt.
    (c) Cumulative block count per refinement level.
    At $t = 1.5$ there are about $10^5$ blocks and $0.4 \times 10^9$ cells.}
  \label{fig:amr-mesh-time}
\end{figure}

The three panels of figure~\ref{fig:amr-mesh-time} show (a) how the mesh changes over time; (b) a snapshot of the solution at $t = 1.5$; and (c) the block counts at the different refinement levels.
The highest refinement level is six, at which the grid spacing is $\Delta x = 1/1024$.
Note that most of the blocks are at level six, even though the other refinement levels cover a comparable amount of volume.
During the simulation, the block count increases from about $1.5 \times 10^4$ to $1.0 \times 10^5$, and the cell count increases from about 62 million to 416 million.
Simulating until $t = 1.5$ takes 39746 iterations, and the time-averaged cell count is about 192 million.
The snapshot shown in figure~\ref{fig:amr-mesh-time-b} corresponds to the solution at the final time.

\begin{figure}
  \centering
  \includegraphics[width=\linewidth]{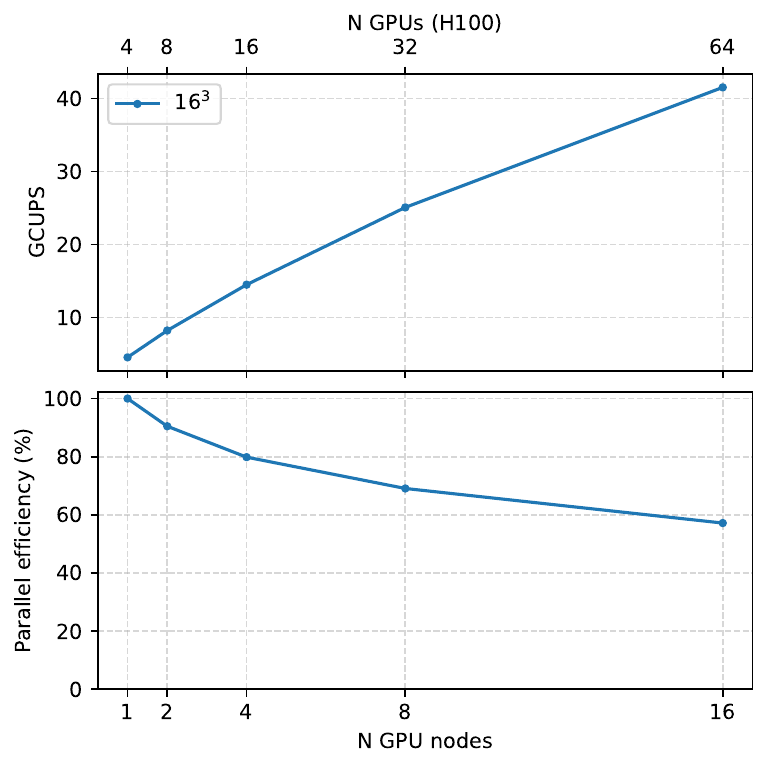}
  \caption{GPU strong scaling results for the Rayleigh--Taylor test case with AMR\jt{ and OpenACC}, using blocks of $16^3$.}
  \label{fig:strong-scaling-amr}
\end{figure}

Figure~\ref{fig:strong-scaling-amr} shows strong scaling results on H100 GPUs.
Since the mesh changes in time, GCUPS values were determined by dividing the total number of cells updated in a simulation by the total runtime.
Using 16 GPU nodes, the parallel efficiency is about \jt{60\%} and a GCUPS value of \jt{41.5} is obtained.
This is lower than the value of 60 for the uniform $512^3$ grid case \jt{with $16^3$ blocks} shown in figure~\ref{fig:strong-scaling-gpu}, but such a difference is to be expected since a case with AMR involves several additional algorithmic steps.

\begin{figure}
  \centering
  \includegraphics[width=\linewidth]{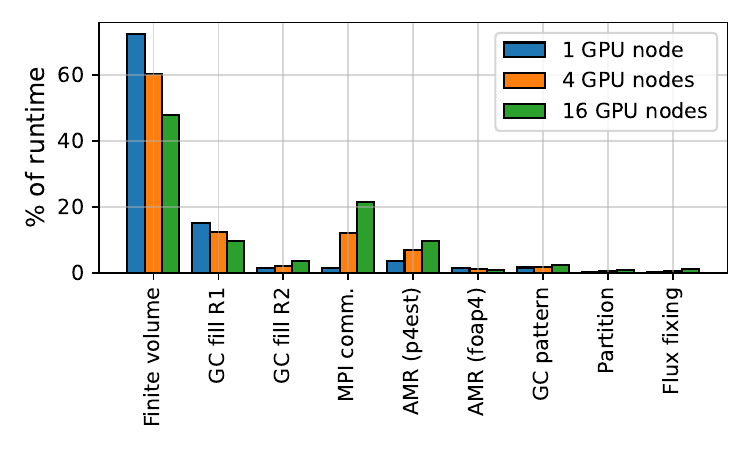}
  \caption{Breakdown of computational cost of foap4 for the Rayleigh--Taylor test case with AMR on \jt{nodes containing four }H100 GPUs. \jt{See the text for a detailed description of the columns. The values given are averages over the MPI ranks.}}
  \label{fig:strong-scaling-amr-barchart}
\end{figure}

A breakdown of \jt{the computational cost} is shown in figure~\ref{fig:strong-scaling-amr-barchart}, \jt{which contains the following columns:
  \begin{itemize}
    \setlength\itemsep{0.0em}
    \item \textbf{Finite volume}: cost of computing fluxes and updating the PDE solution.
    \item \textbf{AMR (p4est)}: cost of updating the mesh structure in p4est, including MPI communication.
    \item \textbf{AMR (foap4)}: cost of adapting the `PDE' mesh (restriction, prolongation of solution)
    \item \textbf{GC pattern}: cost of computing the new stored `pattern' for filling ghost cells on the GPU, see section~\ref{sec:ghost-cell-exchange}.
    \item \textbf{Partition}: cost of re-partitioning the mesh over the MPI ranks
    \item \textbf{GC fill R1/R2}: cost of filling ghost cells on blocks (for the PDE) in round one or two, see section~\ref{sec:ghost-cell-exchange}.
    \item \textbf{MPI comm.}: total cost of all MPI communication of foap4 data, but excluding p4est communication.
    \item \textbf{Flux fixing}: cost of flux fixing, see section~\ref{sec:flux-fixing}.
  \end{itemize}
}
The cost of communication increases with the number of nodes, since the surface-to-volume ratio of each subdomain becomes larger.
The relative cost of updating the mesh in p4est also increases.
This could be related to the parallel communication between CPUs that is required to ensure the mesh remains 2:1 balanced.
The relative cost of filling ghost cells does not change much when the node count increases, and the contributions of other algorithmic components are relatively minor.

\begin{figure}
  \centering
  \includegraphics[width=\linewidth]{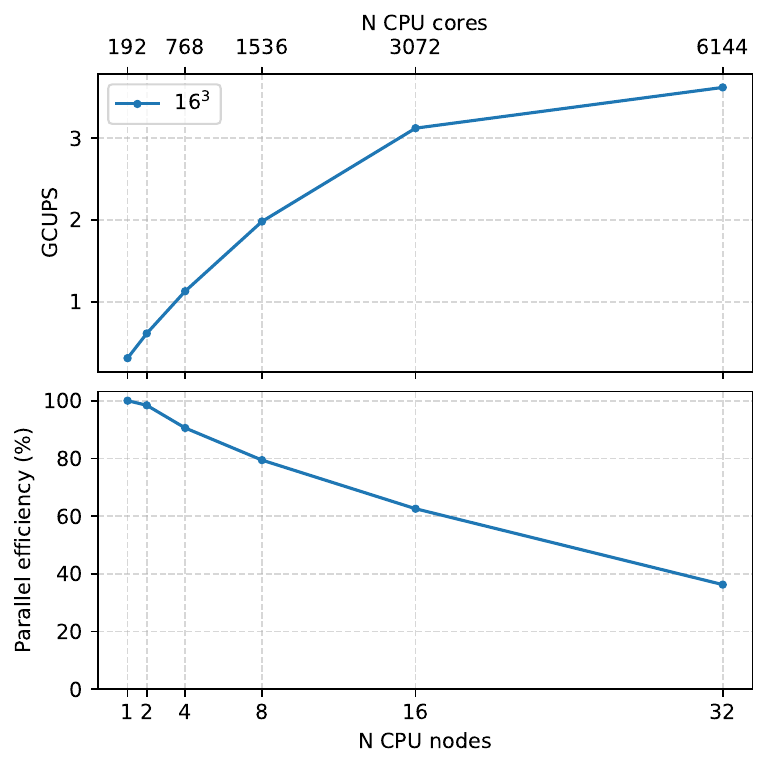}
  \caption{CPU strong scaling results for the Rayleigh--Taylor test case with AMR, using blocks of $16^3$. \jt{Two runs per node count were performed; the best results are shown. Eight OpenMP threads were used per MPI task.}}
  \label{fig:strong-scaling-amr-cpu}
\end{figure}

For comparison, strong scaling results on CPUs are shown in figure~\ref{fig:strong-scaling-amr-cpu}, \jt{using eight OpenMP threads per MPI task.
  The parallel efficiency versus node count is similar on GPU and CPU nodes.}
  GCUPS values are very different however: about \jt{41.5} with 16 GPU nodes and about \jt{$3.6$} with 32 CPU nodes. \jt{At $32$ nodes, the parallel efficiency of the CPU case has already dropped below $40\%$ meaning that the GPU case wins by more than a factor of 10 in terms of minimum time to solution.}


\section{Conclusions}
\label{sec:conclusion}

We have presented foap4, a framework for parallel simulations with adaptive mesh refinement.
\jt{Foap4 is written in Fortran and supports MPI, OpenACC and OpenMP, so that it can run on GPUs but also purely on CPUs.}
The main reason for developing the framework was to have a benchmarking tool for combined MPI \jt{plus OpenACC/OpenMP} parallelization, so that we could make informed decisions on how to add GPU support to existing AMR codes.
For simplicity, the code is therefore limited to 2D or 3D Cartesian grids, with grid blocks of size $N^D$ where $D$ is the problem dimension.
We have demonstrated the performance of foap4 on tests in which Euler's equations of gas dynamics are solved.
These tests range from a 2D uniform grid case on a single GPU to a multi-GPU 3D AMR test case.

The most important conclusion from the paper is that high performance is possible with foap4.
We have demonstrated strong scaling results \jt{for a 3D Rayleigh--Taylor problem.
  On a $512^3$ uniform grid with 64 H100 GPUs, about $65 \times 10^9$ cell updates per second (GCUPS) were obtained with blocks of size $32^3$.
  With blocks of $16^3$ and $8^3$ GCUPS values were about 60 and 50, respectively.
  For a test case with AMR and 64 H100 GPUs about 42 GCUPS was obtained.
  In this test, the number of cells varied between $0.06\times 10^9$ and $0.4\times 10^9$, refinement was performed every four time steps, and blocks of size $16^3$ were used.
}
On GPU nodes, we can get performance that is hard to obtain on CPU nodes, since so many CPU nodes would be necessary that parallel scaling issues become important.
Furthermore, foap4 gives good performance with relatively small block sizes, such as $8^3$ or $16^3$.
Earlier work on GPU AMR simulations often required larger blocks (e.g.\ $64^3$ or $128^3$ in ~\cite{liskaHAMRNewGPUaccelerated2022, stoneAthenaK2026}),
\jt{although good performance with small blocks has recently also been demonstrated in the kalypsso code~\cite{kestenerKalypssoPerformancePortable2026}.}

Even though foap4 was designed for GPUs, the performance on CPUs is still \jt{decent}.
This means that it should be possible to maintain a single code base while supporting both GPUs and CPUs.
Work in this direction is ongoing in the \jt{AGILE project~\cite{porthAstrophysicsGPUsIntroducing2026}}, in which a new \jt{GPU accelerated} version of MPI-AMRVAC is developed.
\jt{The results of the AGILE project} indicate that the design choices tested with foap4 transfer well to MPI-AMRVAC.
However, the foap4 code itself can also be used for real applications, since it already supports different types of boundary conditions, parallel output, and since adding new physics is not difficult.

\section*{Acknowledgements}

This publication was supported by the AGILE project (Grant ID NLESC.OEC.2023.008) funded by the Netherlands eScience Center.
This work used the Dutch national e-infrastructure with the support of the SURF Cooperative using grant no. EINF-11051 ``AGILE''.
This work also made use of resources and expertise provided by SURF Experimental Technologies Platform, which is part of the SURF cooperative in the Netherlands, under project no.~SURF-ETP0037.
J.V. acknowledges support from Research Foundation -- Flanders (FWO) postdoctoral fellowship 1255226N and the KU Leuven postdoctoral mandate PDMT1/24/012. R.K. acknowledges funding from the KU Leuven C1 project C16/24/010 UnderRadioSun and the Research Foundation Flanders FWO project G0B9923N Helioskill.

\section*{Data availability statement}

The foap4 source code is available at \url{https://github.com/jannisteunissen/foap4}

\appendix

\section{\jt{OpenACC with gfortran}}
\label{sec:appendix-gfortran}

\jt{To compile the foap4 code with \texttt{gfortran 14}, a few changes had to be made.
Allocating zero-sized arrays to the device is not supported with this version of \texttt{gfortran}, so we simply use an array size of one in such cases, which otherwise has no effect on code functionality.
Similarly, we do not perform CPU-to-GPU updates over zero elements, and in such cases simply skip the operation.
We also found it necessary to use an associate construct when such update clauses referred to an element of a custom type.
Together, this would result in code as shown below.}
\begin{lstlisting}
associate (array => f4%array)
  if (array_size > 0) then
    !$acc update device(array(1:array_size))
  end if
end associate
\end{lstlisting}
\jt{Another minor change is that module-level scalars that have to be visible on the device were put into user-derived types.}

\jt{We have not yet tried to optimize performance with \texttt{gfortran}, but some initial results are shown in table~\ref{tab:performance-gfortran}.
  The test case is the same 3D Rayleigh-Taylor simulation as for table~\ref{tab:performance-hardware}, but here we consider two limiters: the van Leer limiter, which requires $N_\mathrm{ghost} = 2$, and the WENO5 limiter, which requires $N_\mathrm{ghost} = 3$.
  For the computationally cheap van Leer limiter, GCUPS values are about 35--65\% lower with \texttt{gfortran} than with \texttt{nvfortran}, with the most significant differences for 32-bit floating point numbers.
  With the more expensive WENO5 limiter these differences are considerably smaller, and fp32 performance with \texttt{gfortran} is actually higher on the H100 than with \texttt{nvfortran} and OpenACC.
}

\begin{table*}
  \begin{tabular}{l l l | l l l | l l l}
     & & & \multicolumn{3}{c|}{GCUPS (fp64)} & \multicolumn{3}{c}{GCUPS (fp32)}\\
    Hardware & Compiler & Limiter & $32^3$ & $16^3$ & $8^3$ & $32^3$ & $16^3$ & $8^3$\\
    \hline
    \multirow{3}*{H100} & \texttt{nvfortran-acc} & \multirow{3}*{van Leer} & 1.555 & 1.347 & 1.063 & 3.164 & 2.774 & 2.231\\
    & \texttt{nvfortran-omp} &  & 1.504 & 1.300 & 1.018 & 3.025 & 2.535 & 1.939 \\
     & \texttt{gfortran-acc} &  & 0.859 & 0.829 & 0.696 & 1.373 & 1.326 & 1.233 \\
    \hline
    \multirow{3}*{H100} & \texttt{nvfortran-acc} & \multirow{3}*{WENO5} & 0.679 & 0.612 & 0.471 & 0.935 & 0.862 & 0.743\\
    & \texttt{nvfortran-omp} &  & 0.673 & 0.603 & 0.489 & 1.266 & 1.121 & 0.908\\
     & \texttt{gfortran-acc} &  & 0.501 & 0.467 & 0.399 & 0.954 & 0.895 & 0.779\\
    \hline
    \multirow{2}*{1080 Ti} & \texttt{nvfortran-acc} & \multirow{2}*{van Leer} & 0.099 & 0.096 & 0.087 & 0.410 & 0.378 & 0.254 \\
     & \texttt{gfortran-acc} &  & 0.064 & 0.063 & 0.058 & 0.140 & 0.138 & 0.123 \\
    \hline
    \multirow{2}*{1080 Ti} & \texttt{nvfortran-acc} & \multirow{2}*{WENO5} & 0.034 & 0.033 & 0.031 & 0.145 & 0.138 & 0.118 \\
     & \texttt{gfortran-acc} &  & 0.029 & 0.028 & 0.026 & 0.088 & 0.085 & 0.075 \\
  \hline
\end{tabular}
\caption{Comparison of OpenACC performance with \texttt{gfortran 14.2} and \texttt{nvfortran 25.3} for the Rayleigh--Taylor test case on a uniform $256^3$ grid, with block sizes of $32^3$, $16^3$ and $8^3$.
  For comparison, results with \texttt{nvfortran} and OpenMP offloading are also included.
  For \texttt{gfortran} we used \texttt{-Ofast -march=native -fopenacc -foffload=nvptx-none}, and for \texttt{nvfortran} we used \texttt{-fast} with \texttt{-acc=gpu} or \texttt{-mp=gpu}.
}
\label{tab:performance-gfortran}
\end{table*}

\section{Coefficients for prolongation}
\label{sec:appendix-gminmod}

Below we briefly discuss the maximum parameter $\theta$ that still ensures prolongation with the generalized minmod limiter of equation~\eqref{eq:gminmod} and does not introduce negative values when the original values are non-negative.
To do this, we consider a 1D cell-centered grid with a uniform grid spacing and a `worst-case' scenario, in which the values of three adjacent coarse cells are $U(x-h) = 0$, $U(x) = 1$ and $U(x+h) = M$.
Here $x$ denotes the location of the cell center, $h$ is the grid spacing and $M \gg 1$.
If the grid is refined, there will be two fine cells $u(x-h/4)$ and $u(x+h/4)$ covering the central coarse cell.
We will now apply the generalized minmod limiter given by equation~\eqref{eq:gminmod}, using the slopes $a = 1/h$ and $b = (M-1)/h$.
Since $a < b$ and since they are both positive, the value of $u(x-h/4)$ is then
\begin{equation}
  \label{eq:fine-gminmod}
  u(x-h/4) = 1 - h/4 \min\left[\theta/h, M/(2h)\right].
\end{equation}
In 1D, non-negativity is thus ensured when $\theta \leq 4$, which will also hold for $u(x+h/4)$ by symmetry.
If we consider the same problem in 2D or 3D with coarse values $(0, 1, M)$ along each coordinate, the term with a minus sign in equation~\eqref{eq:fine-gminmod} will appear two or three times, respectively.
Ensuring non-negativity thus requires $\theta \leq 2$ in 2D and $\theta \leq 4/3$ in 3D.

\printcredits

\bibliographystyle{cas-model2-names}

\bibliography{references}

\end{document}